\documentclass[aps,prd,twocolumn,reprint,floatfix,superscriptaddress]{revtex4-2}
\usepackage{tabularx}
\usepackage{graphicx}
\usepackage{amsmath}
\usepackage{amssymb}
\usepackage{comment}
\usepackage{xcolor}
\usepackage{xspace}
\usepackage{multirow}
\usepackage{natbib}
\usepackage{hyperref}
\usepackage{ulem}
\usepackage{dcolumn}
\usepackage{siunitx}
\usepackage{subcaption}
\usepackage{cleveref}
\usepackage{orcidlink}

\newcommand{\nruns}{N_{\rm runs}}
\newcommand{\niter}{N_{\rm iter}}
\newcommand{\algoname}{\texttt{GRITCLEAN}\xspace}
\newcommand{\gtype}[1]{Type-#1}

\DeclareMathOperator*{\argmax}{argmax}

\begin{document}
\title{Glitch veto based on unphysical gravitational wave binary inspiral templates}
\author{Raghav Girgaonkar,\orcidlink{0000-0002-4678-2939}}
\email{raghav.girgaonkar01@utrgv.edu}
\affiliation{Department of Physics and Astronomy, The University of Texas Rio Grande Valley, One West University Blvd.,
Brownsville, Texas 78520, USA}
\author{Soumya D.~Mohanty,\orcidlink{0000-0002-4651-6438}}
\email{soumya.mohanty@utrgv.edu}
\affiliation{Department of Physics and Astronomy, The University of Texas Rio Grande Valley, One West University Blvd.,
Brownsville, Texas 78520, USA}
\affiliation{Department of Physics, IIT Hyderabad, Kandi, Telangana-502284, India}
\begin{abstract}
Transient signals arising from instrumental or environmental factors, commonly referred to as glitches, constitute the predominant background of false alarms in gravitational wave searches with ground-based detectors. Therefore, effective data analysis methods for vetoing glitch-induced false alarms are crucial to enhancing the sensitivity of a search.
We present a veto
method for glitches that impact matched filtering-based searches for
binary inspiral signals. The veto uses unphysical sectors in the space of chirp time parameters  as well as an unphysical extension including negative chirp times to efficiently segregate glitches from gravitational wave signals in data from a single detector. Inhabited predominantly by glitches but nearly depopulated of genuine gravitational wave signals, these unphysical sectors can be efficiently explored using particle swarm optimization. In a test carried out on data taken from both LIGO detectors spanning multiple observation runs, the veto was able to reject $99.9\%$ of glitches with no loss of signals detected with signal-to-noise ratio $\geq 9.0$ and total detector frame mass $\leq 80$~$\textup{M}_\odot$. Our results show that extending a matched filter search to unphysical parts of a signal  parameter space promises to be an effective strategy for mitigating glitches. 
\end{abstract}
\maketitle


\section{Introduction}
\label{sec:intro}
The direct detection of a binary black hole (BBH) merger  by the twin LIGO detectors~\cite{LIGOdescription} in 2015~\cite{PhysRevLett.116.061102} ushered in the new era of gravitational wave (GW) astronomy. Since then, the three detector LIGO-Virgo~\cite{Virgo} network has detected over 90 GW signals from compact binary coalescences (CBCs) across three observing runs (O1, O2~\cite{abbott2019gwtc} and O3~\cite{abbott2021gwtc}). The catalog of detected CBC events includes GW170817~\cite{PhysRevLett.119.161101}, a  binary neutron star (BNS) merger, amongst other BBH and neutron star-black hole (NSBH) mergers. The worldwide network of comparable sensitivity GW detectors has grown with the recent addition of KAGRA~\cite{kagra2019kagra} and it will expand further in the coming years with the planned LIGO-India~\cite{2013IJMPD..2241010U}. The expansion of the network will result in a higher combined sensitivity and an increased rate of CBC detections. 

 Searches for CBC signals~\cite{cannon2021gstlal, Aubin_2021, pycbc}, use the generalized likelihood ratio test (GLRT)~\cite{kay1998fundamentals} in which the log-likelihood ratio of given data is maximized over the space of parameters characterizing the target family of signal waveforms. The GLRT is implemented using cross-correlations between the data and candidate signal waveforms, called templates, in a strategy known as matched filtering~\cite{PhysRevD.44.3819,PhysRevD.49.1707,PhysRevD.49.2658}. While the mathematical form of the GLRT, and its derivative the matched filter, are derived from idealized assumptions of stationarity and Gaussianity of the noise in the output of a GW detector, real data have many features that deviate from these assumptions and cause a degradation in its performance. CBC searches, in particular, incur a higher false alarm rate and hence lower sensitivity, due to the presence of transient signals, also known as \textit{ glitches}, of instrumental or environmental origin~\cite{Abbott_2018}. In some cases, glitches can also overlap with GW signals, as in the case of GW170817, and cause a CBC search pipeline to discard genuine GW signals. Glitches appearing in close proximity to GW signals is not a negligible occurrence: In O3, a total of 18 GW events were affected by the presence of a nearby or overlapping glitch, accounting for $\approx 20 \%$ of the total number of detected GW events~\cite{davis2022subtracting}. Careful glitch-cleaning procedures were required to mitigate the impact of these glitches. 
 
 Although glitches come in a variety of shapes, many of them are observed to fall into distinct classes that correspond to their distinct origins in the hardware or the environment of a detector. The presence of glitch classes has been demonstrated using a variety of machine learning methods such as Deep Convolutional Neural Networks~\cite{Daniel2018}, Support vector machines~\cite{Biswas2013}, t-Sne~\cite{BAHAADINI2018}, random forests~\cite{Biswas2013}, and S-means~\cite{Mukherjee_2010}. The Gravity Spy project~\cite{zevin2017gravity,Glanzer_2023} has provided a high-quality dataset for machine learning-based classifiers and has found about 22 different glitch classes over multiple observing runs of the LIGO detectors~\cite{BAHAADINI2018}. 
 
 Although glitch classification is invaluable for tracking down and eliminating their sources, this has seen limited success. For the majority of glitches present in real GW data, one must supplement the GLRT with
 additional algorithms, called vetoes,  to mitigate the adverse effects of glitches. One such method that has been used widely in different forms in CBC searches is the $\chi^2$-veto~\cite{PhysRevD.71.062001}.
Under a unified formalism~\cite{PhysRevD.83.084002}, a $\chi^2$-veto 
consists of choosing a particular subspace in the null space of a given template, projecting the given data onto this subspace, 
and computing the norm of the projection. By construction, the projection norm should be consistent with that of pure noise if the template happens to match the unknown signal in the data. However, it may be sizeable if the data contains a glitch, thereby serving as a veto statistic. A limitation of this approach is that CBC signals do not constitute a vector space in themselves, in the sense that a linear combination of two CBC signals is not another CBC signal, which makes the null space (and its subspaces) dependent on the unknown parameters of a signal in given data. In addition, computational considerations severely limit the dimensionality of the chosen subspace, which creates a vast number of possible choices for the subspaces.  It has been shown~\cite{nitz2017distinguishing,Choudhary:2022nvs} that the freedom of choosing the subspace can be exploited to maximize projections for specific classes of glitches, resulting in improvements in the detection sensitivity for high-mass CBC signals~\cite{abbott2019gwtc}. However, this could also require retuning the method for different glitch classes or when the classes themselves change due to changes in the detectors and their environment across observation runs.
A different veto strategy is implemented in the GstLAL-based inspiral pipeline~\cite{sachdev2019gstlal} that uses a gating method~\cite{PhysRevD.95.042001} to remove glitches that are strong enough to significantly increase the estimated variance of whitened data from the expected value of unity. 

Additionally, GstLAL also employs a signal-consistency veto in which
the time series of the
matched filter output around its peak value is compared with that expected from a real signal. In~\cite{Ashton_2022}, it was shown that several classes of glitches occupy extreme parts of the parameter space of BBH mergers under matched filtering where no GW signals have been observed, thereby suggesting a possible veto based on mapping the distribution of glitches in this space. 

In addition to the above vetoes that are applied to the output of a search pipeline, several approaches have been explored in which the waveform of a glitch is estimated and subtracted from the data before conducting a GW search. Glitch subtraction using BayesWave~\cite{cornish2015bayeswave}, a wavelet-based method that uses data from a detector network to distinguish between a GW signal and a glitch, was used in parameter estimation for GW170817~\cite{Pankow_GW170817_Glitch_subtract} as well as several subsequent signals~\cite{davis2022subtracting}. Adaptive spline fitting implemented in \texttt{SHAPES}~\cite{mohanty2020adaptive} can fit out different types of glitches~\cite{Mohanty_2023} in the presence of an overlapping CBC signal such as GW170817. 
 Glitschen~\cite{merritt2021transient} uses parametrized waveform models to describe identified glitch classes using principal component analysis. The Antiglitch method\cite{bondarescu2023antiglitch} uses empirical waveform models for the four most commonly occurring classes of short-duration glitches to implement a matched filter for detecting and subtracting them out.  

In this paper we present a novel glitch veto called \algoname (glitch rejection using illegal templates and cross-linking of event attribute numbers) that is based on extending a matched filtering search to unphysical parts of the CBC signal parameter space where the signal waveforms can never arise from a gravitational radiation driven inspiral of two point particles. Similar to the $\chi^2$-veto, it exploits the key idea that a CBC signal is expected to present very different responses when correlated with physical and unphysical CBC templates while a glitch, which does not resemble a CBC signal, may not. However, unlike the $\chi^2$-veto, we do not pick the unphysical waveforms from the null subspace of a candidate
signal or tune them for specific glitch classes.
In addition, besides using the norm of the data projection
on unphysical templates as done in $\chi^2$-vetoes, the method also uses the times of arrival of a signal estimated with physical and unphysical templates.

To create unphysical waveforms, we use the space of chirp time parameters that are obtained by an injective but nonsurjective mapping of the binary component masses. This makes the chirp time space larger than the mass parameter space and, hence, splits it into physical and unphysical regions. Furthermore, the unphysical sector can be significantly enlarged using negative chirp time values, allowing a larger population of glitches to be trapped and vetoed. The segregation of the estimated parameters of glitches and injected GW signals using physical and unphysical templates is seen to be strong enough to veto glitch events in real data across a wide range of signal-to-noise ratios (SNRs). The lack of any assumptions in the method about glitch classes ensures its usefulness irrespective of changes in the makeup of the glitch population from one observation run to another.

Searching over unphysical regions of the parameter space incurs a significant additional computational burden in a conventional matched filtering strategy based on a bank of precomputed waveforms. However, this problem is mitigated in a stochastic search strategy for binary inspiral signals implemented with particle swarm optimization (PSO)~\cite{PSO,weerathunga2017performance,normandin2020towards}. In fact, the inclusion of unphysical regions in
positive chirp time space is an inherent part of such a search that incurs no additional computational cost. This is because PSO works best for hypercubical search spaces while the physical region in positive chirp time space is nonhypercubical. Hence, the search space for PSO must necessarily extend to the bounding hypercube of the physical region and include nonphysical sectors. The search over the negative chirp time space, on the other hand, increases the computational cost of \algoname but not strongly, since it is activated only when a candidate event bypasses the veto in the positive chirp time search.

We tested the performance of \algoname on $\approx 131$ hours of data, free of any observed GW signals, taken from the Hanford (H1) and Livingston (L1) detectors across different observing runs. We used a PSO-based search with templates and
injected signals belonging to the restricted 2PN waveform family~\cite{Blanchet_95}. The 2PN approximation has served well as a workhorse in the development of data analysis methods since it is reliable for binaries in the BNS mass range~\cite{PhysRevD.80.084043}. In this paper, it is extended well beyond its range of 
validity in component masses but this is acceptable for a first demonstration of the central idea that inclusion of unphysical regions of the search parameter space is an effective veto strategy.

Using a detection threshold of ${\rm SNR} = 9.0$,  we found that \algoname was able to reject $99.9\%$ of candidate events, most of them glitches since none were GW signals, that crossed this threshold. The safety of \algoname was tested on 
 signals with total mass $\leq 80$~$\textup{M}_\odot$ over a wide SNR range injected in the same data. It was found that \algoname did not reject any of the injections that passed the detection threshold. It is important to note that this performance is achieved with a single-detector analysis alone. The effectiveness of the veto would increase manyfold when supplemented by coincidence requirements for events detected across multiple detectors. Beyond the above limit on the total mass, the safety of \algoname deteriorates but a reliable assessment requires finding and using unphysical search space regions 
 of more sophisticated waveform approximants \cite{PhysRevD.102.064001,PhysRevD.108.124036,Blanchet_2013}.
 This will be addressed in future work.

The rest of the paper is organized as follows. Section \ref{sec:glrt} contains a brief review of the GLRT formalism used in our analysis, the waveform model used, and the PSO-based search. An outline of our glitch veto scheme is given in Sec. \ref{sec:glitch_veto}, followed by the results in Sec. \ref{sec:results}. Our conclusions and discussions for future work are presented in Sec. \ref{sec:conclusions}.

\section{Signal detection and estimation}
\label{sec:glrt}
In this section, we present a brief overview of the mathematical formalism underlying the GLRT for restricted 2PN signals followed by the implementation details of a PSO-based search.
\subsection{Data and signal models}
\label{data_sig_models}
The data from a GW detector is of the form,
\begin{equation}
    y(t) = s(t) + n(t) \; ,
\end{equation}
where $s(t)$ could be a GW signal or a glitch, and $n(t)$ is  noise. 
In this study, we assume that the GW signal belongs to the restricted 2PN waveform family in which the phase evolution of the signal includes corrections due to GW radiation up to orders of $(v/c)^4$, while the modulation of the amplitude is calculated up to only the Newtonian order~\cite{Blanchet_95}. The spins of the binary components are neglected at this order and the orbit is assumed to have circularized by the time the instantaneous frequency of the signal enters the sensitive 
band of ground-based detectors. At this order, the time evolution of the signal frequency is determined solely by the masses of the binary components. We do not make any assumptions about the signal waveforms of glitches.

The single detector strain response, $h(t)$, is expressed in terms of $h_{+,\times}(t)$, the polarization waveforms in the transverse traceless gauge as follows:
\begin{equation}
    h(t) = F_+(\alpha, \delta, \psi) h_+(t) + F_\times(\alpha, \delta, \psi) h_\times(t) \;.
\end{equation}
Here, $F_+(\alpha, \delta, \psi)$ and $F_\times(\alpha, \delta, \psi)$ are the detector antenna patterns that depend on the sky location of the source, given by the azimuthal angle $\alpha$ and the polar angle $\delta$, and the orientation,  given by the polarization angle $\psi$, of the binary orbit projected on the sky. The GW polarization waveforms, $h_{+,\times}(t)$, can be expressed in the Fourier domain using the stationary phase approximation~\cite{PhysRevD.44.3819} as 
\begin{align}
    \Tilde{h}_{+,\times}(f) &= \int_{-\infty}^\infty dt\, h_{+,\times}(t) e^{-2\pi i f t}\;,\\
    \Tilde{h}_{+}(f) &= \mathcal{A}\frac{1+\cos\iota}{2}f^{-\frac{7}{6}}\exp[-i\Psi(f)] \;, \\
    \Tilde{h}_{\times}(f) &= \mathcal{A}\cos\iota f^{-\frac{7}{6}}\exp[-i(\Psi(f) + \frac{\pi}{2})]\;,
\end{align}
where $\mathcal{A}$ is the overall distance-dependent amplitude, $\iota$ is the inclination angle of the orbital plane of the binary to the line of sight from the detector, and $\widetilde{h}_{+,\times}(f)=0$ for $f\notin [f_\ast, f_{\rm ISCO}]$. 
Here, $f_{\rm ISCO}$ is
the highest instantaneous frequency of the GW signal associated with the inspiral, 
corresponding to the innermost stable circular orbit before plunge, 
and $f_\ast$ is the low-frequency cutoff, assumed to be $30$~Hz in this study, caused by the sharp rise in seismic noise
below which a GW signal cannot be observed.  
Although $f_{\rm ISCO}$ depends on the component masses of a binary system, we simplify our code
implementation by setting $f_{\rm ISCO} = 700$~Hz, which
corresponds to the lowest-mass systems considered in our study. 
The phase $\Psi(f)$ is given by,
\begin{equation}
\label{eqn:phase}
     \Psi(f) = 2\pi f (t_a + \tau)  - \phi_0 - \frac{\pi}{4} + \sum_{j = 0}^{4} \alpha_j \left(\frac{f}{f_*}\right)^{(j-5)/3}\;,
\end{equation}
where the coefficients $\alpha_j$ are defined as,
\begin{align}
    \alpha_0 &= 2\pi f_\ast\frac{3\tau_0}{5}, \\
    \alpha_1 &= 0, \\
    \alpha_2 &= 2\pi f_\ast\tau_1, \\
    \alpha_3 &= -2\pi f_\ast\frac{3\tau_{1.5}}{2}, \\
    \alpha_4 &= 2\pi f_\ast3\tau_2\;,
\end{align}
and $\{\tau_0, \tau_1, \tau_{1.5}, \tau_2\}$, called the {\it chirp time} parameters, 
are obtained from the binary 
component masses $m_1$ and $m_2$ as, 
\begin{align}
    \tau_0 &= \frac{5}{256\pi}f_\ast^{-1} \left(\frac{GM}{c^3}\pi f_\ast\right)^{-\frac{5}{3}} \eta^{-1}, \\
    \label{eq:tau1}
    \tau_1 &= \frac{5}{192\pi}f_\ast^{-1} \left(\frac{GM}{c^3}\pi f_\ast\right)^{-1} \eta^{-1}\left(\frac{743}{336} + \frac{11}{4}\eta \right), \\
    \tau_{1.5} &= \frac{1}{8}f_\ast^{-1} \left(\frac{GM}{c^3}\pi f_\ast\right)^{-\frac{2}{3}} \eta^{-1}, \\
    \label{eq:tau2}
    \tau_2 &= \frac{5}{128\pi}f_\ast^{-1} \left(\frac{GM}{c^3}\pi f_\ast\right)^{-\frac{1}{3}} \\ &\times\eta^{-1}\left(\frac{3058673}{1016064} + \frac{5429}{1008}\eta + \frac{617}{144}\eta^2 \right) \;,\\
    \label{eq:massterms}
    M &= (m_1 + m_2),\: \mu = \frac{m_1m_2}{M}, \: \eta = \frac{\mu}{M} \;.
\end{align}
The time at which the instantaneous frequency of the
signal crosses $f_\ast$ is designated as its time of arrival (TOA) $t_a$ and the phase of the signal at $t=t_a$ is the initial phase $\phi_0$. The combination 
\begin{equation}
\label{eq:chirplengtheqn}
    \tau = \tau_0 - \tau_{1.5} + \tau_1 + \tau_2\;,
\end{equation}
called the \textit{chirp length}, approximates the observed 
duration of the signal  and is a key quantity in \algoname: For a physically valid waveform, one must have $\tau > 0$.
The restricted 2PN signals can be parametrized either in terms of the binary component masses or any two of the chirp times. It is convenient to choose $\tau_0$ and $\tau_{1.5}$ as the independent parameters since they can be 
inverted analytically to obtain $M$ and $\mu$, hence $m_1$ and $m_2$,
\begin{align}
\label{eq:derivedmassterms_1}
    \mu &= \frac{1}{16f_*^2}\left( \frac{5}{4\pi^4\tau_0 \tau_{1.5}^2} \right)^{\frac{1}{3}} \left(\frac{G}{c^3}\right)^{-1},\\
    \label{eq:derivedmassterms_2}
    M &= \frac{5}{32f_*}\left( \frac{\tau_{1.5}}{\pi^2\tau_0} \right) \left(\frac{G}{c^3}\right)^{-1}\;,\\
\label{eq:est_m1} 
    m_1 &= \frac{M - \sqrt{M^2 - 4 \mu M}}{2} \;, \\
    \label{eq:est_m2}
    m_2 &= \frac{M + \sqrt{M^2 - 4 \mu M}}{2}\;,
\end{align}
where we choose the convention $m_1 < m_2$.
The values of $M$ and $\mu$ above can in turn be used to derive the remaining chirp times $\tau_1$ and $\tau_2$. 
\subsection{The GLRT and MLE}
While not true in general for real data, the assumption of $n(t)$ being a Gaussian noise process allows us to derive the log-likelihood ratio of a finite duration segment of the data as,
\begin{eqnarray}
    {\rm LLR}(\overline{y}, \Theta ) & = & \langle \overline{y}, \overline{s}(\Theta) \rangle - \frac{1}{2} ||\overline{s}(\Theta)||^2\;,
\end{eqnarray}
where for any continuous time function $x(t)$, $\overline{x} = (x_0, x_1, \ldots, x_{N-1})$ is the row vector of sampled values $x_k = x(t_k)$, with $t_k = k/f_s$ and $f_s$ being the sampling frequency, and
\begin{eqnarray}
    \langle \overline{x},\overline{y}\rangle & = & \overline{x}{\bf C}^{-1}\overline{y}^T\;,
\end{eqnarray}
is the inner product based on the covariance matrix ${\bf C}$ of the noise $\overline{n}$ in $\overline{y}$, with $C_{ij} = {\rm E}[n_i, n_j]$, and   
$\|\overline{x}\|^2  =  \langle \overline{x},\overline{x}\rangle$ is the norm induced by the inner product.
 Here, $\Theta$ is the set of parameters that describe the 2PN signal. 
 Under the assumption of wide-sense stationarity of the noise, the inner product above is conveniently 
 expressed as 
\begin{eqnarray}
    \langle \overline{x}, \overline{y}\rangle & = & \widetilde{x} \left(\widetilde{y}./ \overline{S}_n\right)^\dagger \;,
\end{eqnarray}
where $\widetilde{x}$ is the discrete fourier transform (DFT) of $\overline{x}$, `$./$' denotes element-wise division, and $\overline{S}_n$ is the  two-sided power spectral density (PSD) $S_n(f)$ sampled at the DFT frequencies. 

In GLRT, the detection statistic is defined as,
\begin{eqnarray}
    \label{eqn:glrt}
    L_G (\overline{y}) & = & \max \limits_{\Theta} {\rm LLR}(\overline{y}, \Theta )\;.
\end{eqnarray}
Formally, a signal detection is declared if $L_G(\overline{y}) > \eta$ for a preset threshold $\eta$. In such a case,
\begin{eqnarray}
    \widehat{\Theta} & = & \argmax\limits_{\Theta} {\rm LLR}(\overline{y},\Theta)\;,
\end{eqnarray}
provides the maximum likelihood estimate (MLE) of the unknown parameters of the signal present in the data. Henceforth, we will call the pair $(L_G(\overline{y}),\widehat{\Theta})$ for a given data segment $\overline{y}$ an {\it event}.
In real data, which may have non-GW signals or non-Gaussian noise, the result of 
the GLRT is not immediately deemed to be a detection even if $L_G(\overline{y})> \eta$. 
Instead, the occurence of $L_G(\overline{y})> \eta$ provides only a {\it candidate  event} for the given $\overline{y}$ that must pass additional
checks, namely, the vetoes before it can be declared to be a GW detection.

For the single detector case considered in this paper, $F_{+,\times}$, $\iota$, and $\mathcal{A}$ fold into an overall amplitude and a redefinition of the initial phase $\phi_0$. Splitting the set of parameters as $\Theta = \Theta^\prime \cup \{\rho,\phi_0\}$, where $\Theta^\prime = \{\tau_0, \tau_{1.5},t_a\}$ and $\rho = \|\overline{s}(\Theta)\|$,  the signal can be expressed as,
\begin{eqnarray}
    \overline{s}(\Theta) & = & \rho \overline{q}(\Theta^\prime,\phi_0)\;,\\
    \overline{q}(\Theta^\prime,\phi_0) & = & \frac{\overline{s}(\Theta)}{\|\overline{s}(\Theta)\|}\;,\Rightarrow \|\overline{q}(\Theta^\prime,\phi_0)\| = 1\;.
\end{eqnarray}
Here, $\rho$  defines
the SNR  $\overline{s}(\Theta)$ and $\overline{q}(\Theta^\prime,\phi_0)$ is called 
the template waveform. 

In terms of these parameters, the GLRT can be computed as,
\begin{align}
    L_G(\overline{y}) &= \max \limits_{\Theta^\prime}\max \limits_{\rho,\phi_0} \left( \rho \langle \overline{y}, \overline{q}(\Theta^{\prime},\phi_0) \rangle - \frac{1}{2} \rho^2  \right) \;. \label{eq:L_G_rho_phi0_Thetap}
\end{align}
Substituting the solution of the inner maximization, namely,
\begin{eqnarray}
    \widehat{\rho} &=&   \langle \overline{y}, \overline{q}(\Theta^\prime,\widehat{\phi}_0) \rangle \;, \label{eq:est_rho}\\
    \widehat{\phi}_0 & = & \arctan{\frac{\langle \overline{y}, \overline{q}_0(\Theta^{\prime}) \rangle}{\langle \overline{y}, \overline{q}_1(\Theta^{\prime}) \rangle}}\;,\label{eq:est_initphase}\\
    \overline{q}_0(\Theta^\prime) & = & \overline{q}(\Theta^\prime, 0)\;,\\
    \overline{q}_1(\Theta^\prime) & = & \overline{q}(\Theta^\prime, \pi/2)\;,
\end{eqnarray}
in Eq.~(\ref{eq:L_G_rho_phi0_Thetap}) and using another split, $\Theta^\prime = \theta \cup \{t_a\}$ and $\theta = \{\tau_0,\tau_{1.5}\}$,
the expression for the GLRT statistic reduces to 
\begin{eqnarray}
    L_G (\overline{y}) &=& \max \limits_{\theta} \lambda (\theta)\;,\label{eq:Lg_over_theta}\\
    \lambda (\theta)   &=& \max \limits_{t_a} \left[ \langle \overline{y}, \overline{q}_0(\theta,t_a) \rangle^2  + \right. \nonumber\\
                    &&\left. \langle \overline{y}, \overline{q}_1(\theta,t_a) \rangle^2 \right) ]\;.
\end{eqnarray}
For fixed $\theta$, the maximization over $t_a$ can be performed by simply evaluating $\langle \overline{y}, \overline{q}_{0,1}(\theta, t_a)\rangle$ at values
of $t_a$ equal to the sampling times $t_k = k/f_s$, $k = 0, 1,\ldots,N-1$. This is a cross-correlation operation between the data and each of the quadrature
templates, $\overline{q}_0(\Theta^\prime)$ and $\overline{q}_1(\Theta^\prime)$, which can be performed
efficiently~\cite{5217220,schutz_1991} using the FFT.
In this paper, the remaining maximization of $\lambda(\theta)$, called the \textit{fitness function}, over $\theta$ is carried out using PSO. 

Note that, from Eq.~(\ref{eq:est_rho}) and Eq.~(\ref{eq:est_initphase}), the square root of $L_G(\overline{y})$ is equivalent to the SNR of the estimated signal. As such, 
from here on, we refer to $\sqrt{\lambda(\theta)}$ as the {\it estimated} SNR at the location $\theta$.

\subsection{PSO-based search}
\label{sec:pso}
In the specific context of the optimization problem in Eq.~(\ref{eq:Lg_over_theta}),
PSO searches for the global maximum of the fitness function $\lambda(\theta)$ iteratively 
over a rectangle $\tau_{a,{\rm min}}\leq \tau_a\leq \tau_{a,{\rm max}}$, $a\in {0, 1.5}$ called the {\it search space}. In each 
iteration, a fixed number of sampled values are obtained for the fitness 
function. The locations of these samples in the search space are called {\it particles} and the set of particles is called a {\it swarm}. Based on their fitness values, the particles are moved to new locations 
in the next iteration. Each particle keeps track of the best location in its own 
history, called its {\it personal best}, as well as the best location found by its neighbors, called its {\it local best}. The specification 
of the neighborhood of each particle sets the topology of the swarm. Each 
particle carries with it a vector, called its {\it velocity}, that specifies its displacement from one iteration to the next. 

The heart of the PSO algorithm is the dynamical equation  for updating the velocity. In standard variants of PSO, it has contributions from (i) an
{\it inertia} term that simply scales the current velocity, (ii) a randomly weighted 
{\it cognitive term} that attracts the particle towards its personal best location, and (iii) a randomly weighted {\it social term} that attracts the particle towards the local best location. The final result of the search is the  best of all local bests at termination. A simple termination condition, followed in our work, is to  fix
the number of iterations, $\niter$, in advance. One also needs to specify the dynamical rule for  a particle that breaches the search space boundary. We use the let-them-fly boundary condition~\cite{bratton2007defining} in which no change is made to the dynamical equation but the fitness is set to $-\infty$, thereby ensuring that the attractions of the local and personal bests eventually pull the particle back in.  
Further details about PSO are discussed in  pedagogical introductions such as ~\cite{robinson2004particle,mohanty2018swarm}.

A stochastic search algorithm, such as PSO, is not guaranteed to converge to the global maximum in a finite number of iterations. However, the probability of convergence can be improved by tuning the algorithm for a specific problem. Further improvement can be obtained by using multiple runs  of the algorithm, with independent pseudorandom streams, on the same fitness function and
selecting the best solution among them~\cite{mohanty2018swarm}. PSO stands out as a particularly robust algorithm, at least in its applications to GW data analysis problems, and the only parameters that typically need to be tuned are $\niter$ and the number, $\nruns$, of independent runs. 

\begin{figure}
    \centering
    \includegraphics[scale = 0.16]{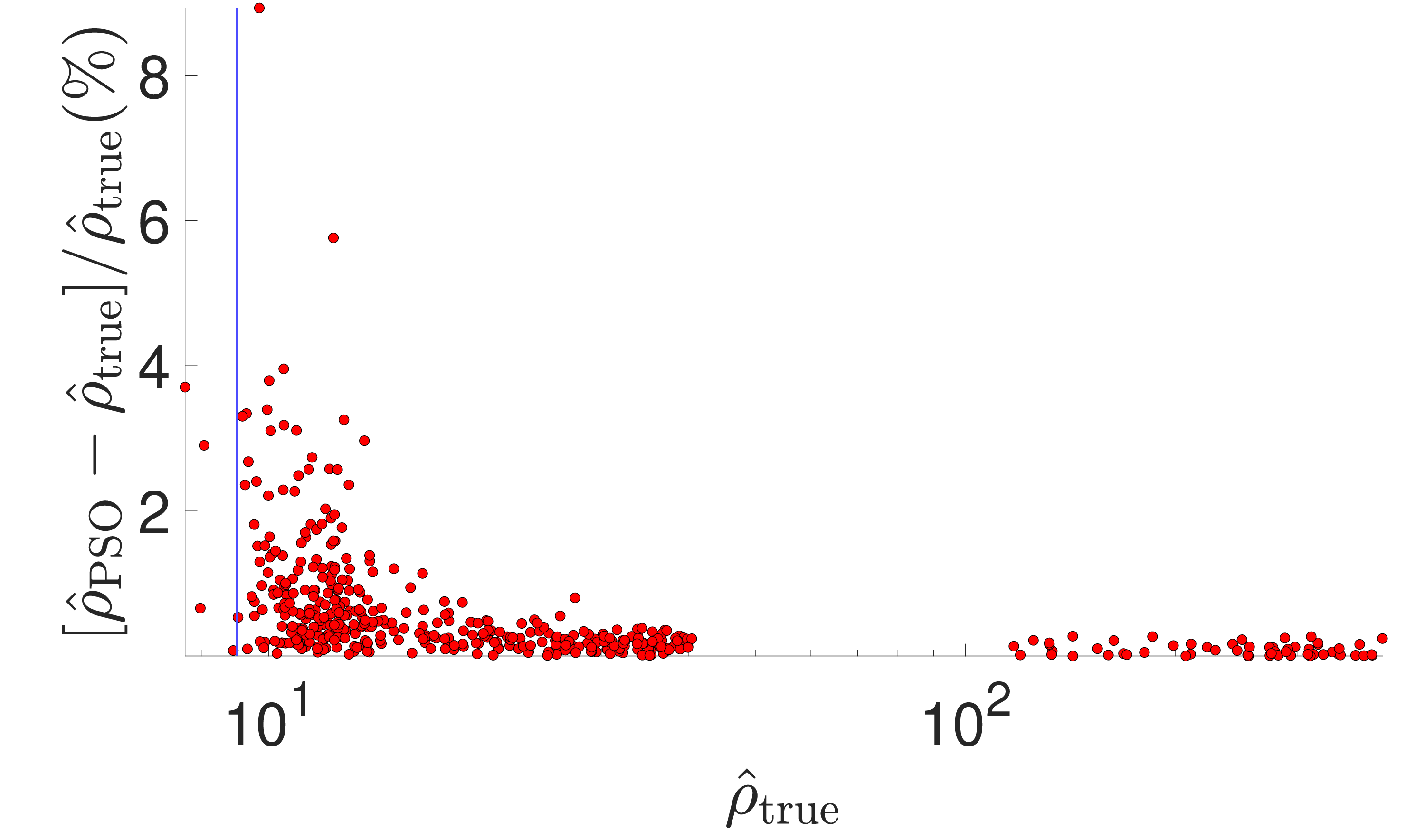}
    \caption{Plot illustrating performance of PSO. Each point corresponds to one data segment containing an injected GW signal. Here, $\widehat{\rho}_{\rm true}$  and $\widehat{\rho}_{\rm PSO}$ denote the estimated SNRs at the true parameters of the injected signal and the best location found by PSO, respectively. The y-axis shows the relative difference in these two quantities, with the vertical blue line showing ${\rm SNR}=9.0$ for reference. In all cases,  $\widehat{\rho}_{\rm PSO} > \widehat{\rho}_{\rm true}$, indicating that PSO performed well.
    }
    \label{fig:psoperf}
\end{figure}

A simple strategy to assess the tuning of PSO is to take a data realization with an injected GW signal and compare the estimated SNR found by PSO, $\widehat{\rho}_{\rm PSO}$,  with the value, $\widehat{\rho}_{\rm true}$, at the true parameters of the injection.   Since, in the presence of noise, the global maximum of the fitness function must shift away from the injection parameters, one expects to have $\widehat{\rho}_{\rm PSO}\geq \widehat{\rho}_{\rm true}$ if PSO is working well. A given choice of $\nruns$ and $\niter$ is deemed acceptable if this condition is satisfied in a  significant fraction of independent data realizations. In this paper, we have set  $N_{\rm iter} = 500$ and $N_{\rm run} = 8$, and as shown in Fig.~\ref{fig:psoperf},  the condition $\widehat{\rho}_{\rm PSO}\geq \widehat{\rho}_{\rm true}$ is
satisfied in all the data segments with injected signals (see Sec.~\ref{sec:siginj})
used in our study. The other parameters of PSO have been set at the same values as in~\cite{normandin2020towards}. In particular, we use $40$ particles and the ring topology where the neighborhood of each particle consists of the two particles adjacent to it when the indices of the particles are placed sequentially around a circle.

PSO is known to perform best when the search space is a hypercube. Nonhypercubical search spaces lead to an excessive loss of particles under the boundary condition used here. Thus, when run over the rectangular chirp time search space described above, the search naturally extends to its unphysical sectors. The PSO-based search was carried out using a pipeline that was developed in \texttt{MATLAB R2022}~\cite{MATLAB:R2022a_u7}. For each data segment in our study, the search ranges for PSO were set as $\tau_0 \in [0,90]$, $\tau_{1.5} \in [0,2]$ for the positive chirp time quadrant, and $\tau_0 \in [-90,0]$, $\tau_{1.5} \in [-2,0]$ for the negative chirp time quadrant.  This is not possible in a search over the mass parameter space, since every point corresponds to a physical signal waveform. We use this property of the chirp time space in our veto scheme as described in the following section.

\section{Glitch veto}
\label{sec:glitch_veto}
As outlined earlier, \algoname uses physical and unphysical template waveforms to probe the nature of a candidate event.
In this section, we first define physical and unphysical templates, and then present the veto strategy. While the negative quadrant of the chirp time space, $\tau_0<0$, $\tau_{1,5}<0$, denoted as $S_N$, consists entirely of unphysical templates, the positive quadrant, $\tau_0>0$, $\tau_{1.5}>0$,  denoted as $S_P$, has both types of templates. At the same time, a part of $S_N$ must be treated specially because the templates in that part can resemble binary inspiral signals despite being unphysical. These details are presented separately for the two quadrants below. 

\subsection{Positive chirp time quadrant}
The quadrant $S_P$, has two different types of unphysical sectors. 
The first is one in which the inversion of $(\tau_0, \tau_{1.5})$ to the masses of binary components $(m_1, m_2)$ yields complex values.
From Eq.~(\ref{eq:est_m1}) and Eq.~(\ref{eq:est_m2}), this happens when
\begin{equation}
    \label{eqn:imaginaryeqn}
    M < 4\mu\;.
\end{equation} 
Expressing this condition in terms of the chirp times using Eqs. (\ref{eq:derivedmassterms_1}) and (\ref{eq:derivedmassterms_2}), we see that points in the
$(\tau_0, \tau_{1.5})$ space that lie below
the curve 
\begin{equation}
    \tau_{1.5} = \left[\frac{\tau_0^2}{f_*^3}\left( \frac{128\pi^2}{25}\right)\right]^{1/5}\;,
\end{equation}
which corresponds to the $M = 4\mu$ or $m_1 = m_2$ curve, satisfy this condition.
The second unphysical region in $S_P$ is the one where the chirp length of a signal, given by Eq.~(\ref{eq:chirplengtheqn}), becomes negative.
This happens due to the fact that $\tau_{1.5}$ appears with an opposite sign to all the other chirp times in the expression for the chirp length and, in this region, $\tau_{1.5}$ acquires a comparatively large value. Note that if $m_1$ is real it must also be positive since for it to be real one must have $M^2 -4\mu M \geq 0$, which necessarily implies that $\sqrt{M^2 -4\mu M} < M$ and that the numerator of Eq.~(\ref{eq:est_m1}) is positive. [For $m_2$, the numerator of Eq.~(\ref{eq:est_m2}) is always positive if it is real.]

\begin{figure*}[t] 
    \centering
    \includegraphics[scale = 0.15]{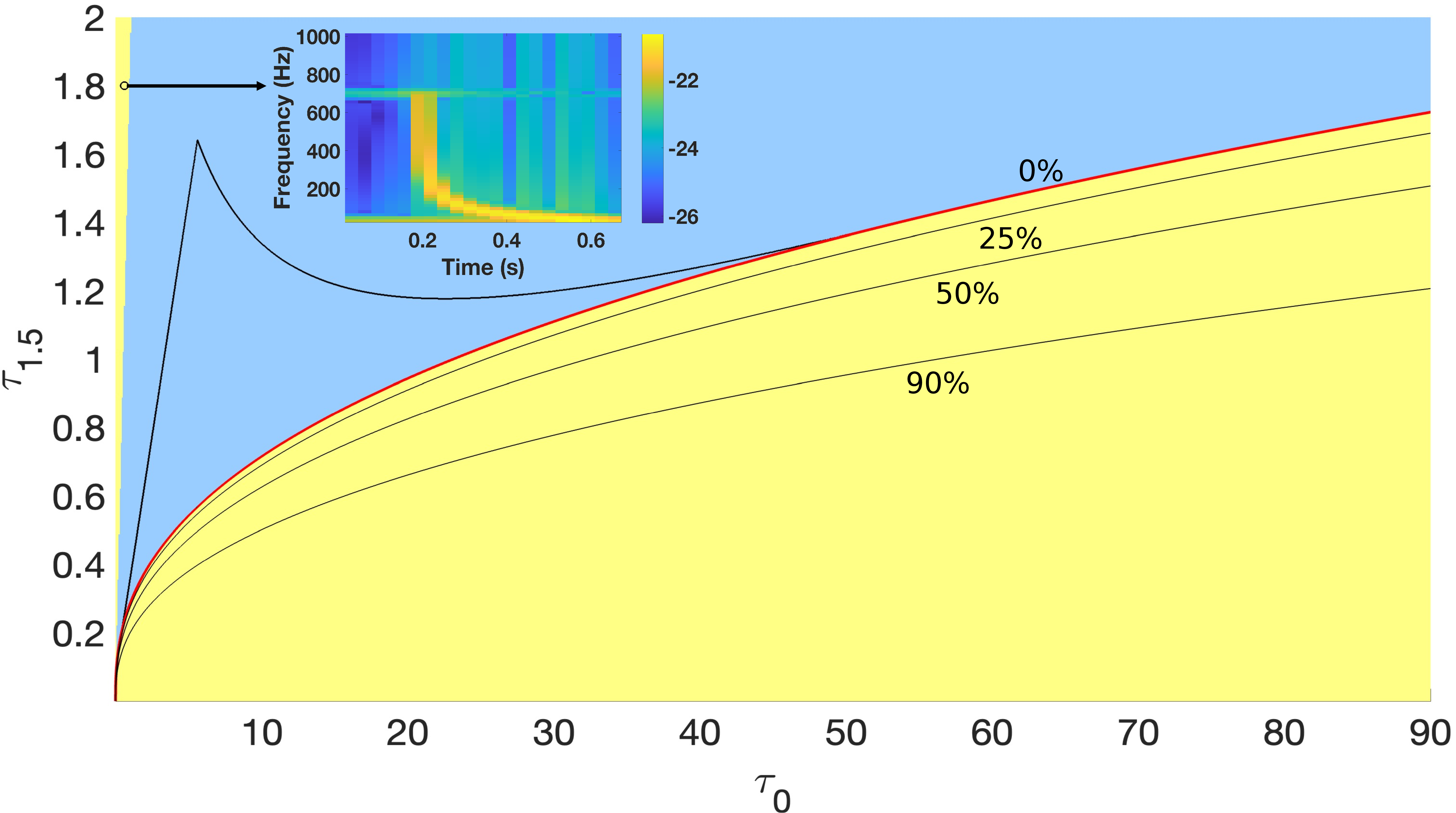}
    \caption{The physical (blue) and unphysical (yellow) regions in the positive chirp time quadrant $S_P$. The wedge-shaped unphysical region on the left corresponds to negative chirp lengths. The decreasing instantaneous frequency of signals in this sector is illustrated in the logarithm of the spectrogram, plotted in arbitrary units, of an example chirp corresponding to a marked point of $[\tau_0,\tau_{1.5}] = [0.6,1.8]$~sec shown as an inset. The remaining unphysical region corresponds to masses with imaginary components. The black curves denote the boundaries for different values of $\zeta$ 
    [Eq.~(\ref{eq:imagratioexpression})]. The kite-shaped boundary corresponds to component masses in the range $[1.4,30] \textup{M}_{\odot}$.
    \label{fig:postauspace}}
\end{figure*}

The physical and unphysical regions of $S_P$ are shown in Fig.~\ref{fig:postauspace}. While the unphysical region corresponding to negative chirp length 
is a very small in area, we find it to be an important one since many glitches are observed to fall in this region while injected signals never do. The template waveforms for this sector are chirps with instantaneous frequencies that decrease with time, as illustrated in Fig. \ref{fig:postauspace}, instead of increasing like a normal GW radiation-driven inspiral.

\subsection{Negative chirp time quadrant}
While the chirp time parameters are never negative by construction, nothing prevents us from using negative values for them in the phase $\Psi(f)$ in Eq.~(\ref{eqn:phase}). The resulting waveform
in the time domain does not correspond to any physical binary inspiral but remains a valid real-valued function of time in itself. Therefore, extending a PSO-based search to $S_N$ provides a significant enlargement of the space of unphysical template waveforms for use in \algoname. At the same time, no additional ad hoc parameters are introduced in the overall veto strategy, such as assumptions about specific glitch classes, besides the search range in $S_N$.  

We define the template waveforms in $S_N$ using a few modifications to the expressions in Sec.~\ref{data_sig_models}. To calculate the values of $\tau_1$ and $\tau_2$ from negative values of $\tau_0$ and $\tau_{1.5}$,  $M$ and $\mu$ are obtained as,
\begin{align}
\label{eq:new_mu}
    \mu &= \frac{1}{16f_\ast^2}\left( \frac{5}{4\pi^4 |\tau_0| |\tau_{1.5}|^2} \right)^{\frac{1}{3}} \left(\frac{G}{c^3}\right)^{-1}\;,\\
    \label{eq:new_M}
    M &= \frac{5}{32f_\ast}\left( \frac{|\tau_{1.5}|}{\pi^2 |\tau_0|} \right) \left(\frac{G}{c^3}\right)^{-1}\;.
\end{align}
The chirp times $\tau_1$ and $\tau_2$ are obtained by using the above expressions for $M$ and $\mu$  in Eq. (\ref{eq:tau1}) and Eq. (\ref{eq:tau2}), respectively, and multiplying both 
by $-1$, making all the chirp time parameters negative.
In addition, we modify $\Psi(f)$ to, 
\begin{equation}
    \Psi(f) = 2\pi f t_a - \phi_0 - \frac{\pi}{4} + \sum_{j = 0}^{4} \alpha_j \left(\frac{f}{f_*}\right)^{(j-5)/3}
\end{equation}
to ensure that the TOA, $t_a$, refers to the start of the waveform.  

\begin{figure*}[t] 
    \centering
    \includegraphics[scale = 0.25]{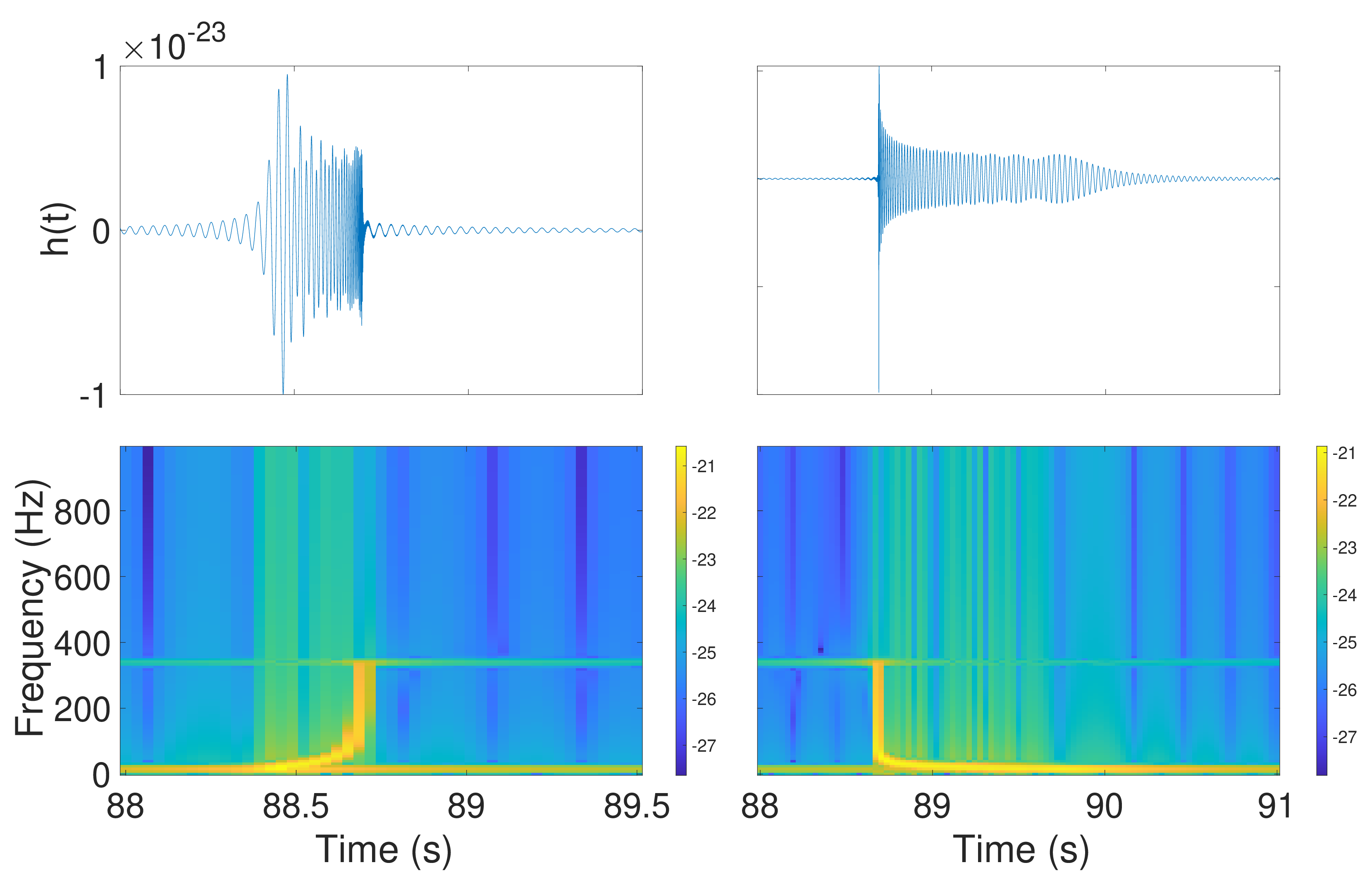}
    \caption{(Left column) An increasing frequency chirp signal in the negative chirp time quadrant $S_N$ with chirp length $\tau > 0$ with the top and bottom panels showing the time series and spectrogram, respectively. The chirp times for this signal are $[\tau_0, \tau_{1.5}] = [-0.6,-1.5]$. 
    (Right column) The time series (top) and spectrogram (bottom)
    of the signal obtained by swapping the values of $\tau_0$ and $\tau_{1.5}$. This
    yields a decreasing frequency chirp.}
    \label{fig:negtauspace_increasingchirp}
\end{figure*}

Almost all of the template waveforms thus defined have negative chirp lengths [cf., Eq.~(\ref{eq:chirplengtheqn})] and instantaneous frequencies that
decrease with time.
However, we find that there exists a small region in $S_N$, with an origin similar to that of the negative chirp time region in $S_P$, in which the waveforms have positive chirp lengths and increasing instantaneous frequencies. Such waveforms can have a strong correlation with GW signals, especially with high-mass and short-duration ones. We address this issue simply by replacing the fitness value for a point $(\tau_0,\tau_{1.5})$ in this region with the value
of the point with the two (still negative) chirp times swapped, that is, $(\tau_{1.5},\tau_0)$. The degeneracy introduced
in the fitness function by this swap
does not have a significant impact on the search in $S_N$ due to the smallness of the region over which this occurs.
Figure \ref{fig:negtauspace_increasingchirp} illustrates the effect of  swapping the chirp times of an increasing frequency chirp in
$S_N$. 

\subsection{Veto strategy}
\label{sec:veto_strategy}

For any candidate event found in the positive quadrant ($S_P$) search,
the application of \algoname involves a sequence
of three veto steps as described below.

\subsubsection{\textbf{Chirp length veto}} 
In the first step, we use the estimated chirp times  $(\tau_0, \tau_{1.5})$ to calculate the chirp length $\tau$ using Eq.~(\ref{eq:chirplengtheqn}). If the chirp length is negative, the candidate is vetoed outright as a glitch. This hard criterion is motivated by our observation that an injected GW signal, regardless of its SNR or other parameters,
never appears in this small sector of $S_P$, while it is well populated with glitches. 

\subsubsection{\textbf{Complex mass veto}} If $\tau > 0$, the next veto step consists of examining the estimated binary component masses. 
Recall that a significant part of $S_P$ corresponds to complex masses.  
 Although a CBC signal will never have chirp time parameters in this unphysical sector, the global maximum of the fitness function and the corresponding estimated signal parameters may fall in this region because of the presence of noise in the data. Therefore, in order to reduce the chances of falsely dismissing a GW signal, one should not reject outright a candidate that falls in this unphysical region but adopt a softer criterion. Accordingly, in the second veto step, we reject a candidate only if the ratio, $\zeta$, of the imaginary to real part in the inferred masses
\begin{equation}
    \zeta = \sqrt{\Bigl \lvert 1 - \frac{4 \mu}{M} \Bigr \rvert} \; ,
\end{equation}
crosses a preset threshold.
In terms of chirp times, this is expressed as
\begin{equation}
\label{eq:imagratioexpression}
    \zeta =  \sqrt{\Biggl \lvert 1 - \frac{8}{5f_*} \left( \frac{5 \pi^2 \tau_0^2}{4 \tau_{1.5}^5} \right)^{1/3} \Biggr \rvert}\;,
\end{equation}
with the expression for the boundary for a particular value of $\zeta$ written as
\begin{equation}
    \label{eq:ratioboundary}
    \tau_{1.5} = \left[ \frac{5 \pi^2 \tau_0^2}{4} \left( \frac{5 f_\ast}{8} (1 + \zeta^2) \right)^{-3} \right]^{1/5} \;.
\end{equation}
The corresponding curves for different values of $\zeta$ are shown in Fig.~\ref{fig:postauspace}.
\subsubsection{$\boldmath{S_N}$ \textbf{veto}}
For any event which bypasses the two previous vetoes,
a second search is launched for the global maximum 
of $L_G(\overline{y})$ in the negative chirp time quadrant $S_N$. This furnishes a new set of estimated parameters, $\widehat{\Theta}_N$, from the $S_N$ search, with $\widehat{\Theta}_P$ being estimated from the $S_P$ run, for a given data segment. Of these parameters, we use the estimated TOA, $\widehat{t}_{a,P}$, and $\widehat{t}_{a,N}$, and the SNR,
$\widehat{\rho}_P$, and $\widehat{\rho}_N$, to construct the TOA difference magnitude
\begin{align}
    |\Delta t_a| &= |\widehat{t}_{a,P} - \widehat{t}_{a,N}| \; , 
    \label{eq:delta_ta_def}
\end{align}
and the magnitude of the relative SNR difference
\begin{align}
    |\Delta \rho| &= \frac{|\widehat{\rho}_{P} - \widehat{\rho}_{N}|}{\widehat{\rho}_{P}}\;.
    \label{eq:delta_rho_def}
\end{align}
A candidate signal is vetoed in \algoname if it falls in a predefined area 
of the $(|\Delta\rho|,|\Delta t_a|)$ plane. The area is 
obtained, as described in Sec.~\ref{sec:results}, using the observed distribution of injected signals in the $(|\Delta\rho|,|\Delta t_a|)$ plane. 

The intuitive motivation for using $|\Delta t_a|$ and $|\Delta \rho|$ is that a GW signal is not expected to correlate well with any template in $S_N$ since the latter are all decreasing frequency chirps. This should lead to a large difference in the estimated SNRs as well as other parameters. However, the only other parameter
that is well-estimated and that can be compared across the searches in $S_P$ and $S_N$ is the TOA since the estimated chirp times have opposite signs, leading to real and complex masses in the two quadrants, while the initial phase is a poorly estimated parameter subject to large errors. For a glitch, on the other hand, that does not closely resemble any of the binary inspiral templates in the physical sector of $S_P$, one expects no particular preference for a high
correlation in either $S_P$ or $S_N$. Therefore, the gap between the estimated SNR and TOA for a glitch is likely to be small. As we shall see in Sec.~\ref{sec:results} our results bear out this intuitive argument quite well. 

\section{Results}
\label{sec:results}
We present here our main results on the performance of \algoname. This is 
preceded by a description of 
the dataset used in our study, 
the preconditioning applied to the data, and the GW signals that were injected to study the safety of \algoname. 
\subsection{Data description}
\label{sec:data_describe}
\begin{table}
    \centering
    \begin{tabular}{c  c  c  c}
    \hline
    \hline
    Observing & Start  & End & Detector \\
    run & GPS & GPS & \\
    & (sec) & (sec)  & \\
    \hline
     O1 & 1127002112 & 1127047168 & H1 \\
     & & & \\
     O2 & 1171222528 & 1171263488 & L1 \\
     & & & \\
    \multirow{2}{*}{O3a} & 1243394048 & 1243508736 & H1 \\
    & 1246453760 & 1246543872 & L1 \\
    & & & \\
     \multirow{2}{*}{O3b} & 1258631168 & 1258749952 & H1 \\
    & 1260441600 & 1260539904 & L1 \\
    \hline
    \hline
\end{tabular}
    \caption{The detectors, the GPS start and end times, and the observing runs of the data used in this study. 
    }
    \label{tab:data}
\end{table}

For this study, we used single detector LIGO data from the Gravitational Wave Open Science Center (GWOSC)~\cite{Abbott_2023}. A total of  $\approx 141$~hours of data was used, distributed across all available observing runs (O1, O2, O3a and O3b), from the Hanford (H1) and Livingston (L1) detectors. All of the selected data passes the $\texttt{CBC\_CAT3}$ flag, which means that the detectors were operating in science mode. More details about the data, namely, the GPS times, the observing runs, and detectors are summarized in Table \ref{tab:data}. By using data from all of the observing runs and the two LIGO detectors, we 
were able to test the performance of \algoname across a broad cross section of the observed glitch population. 

The PSO-based GLRT, $L_G(\overline{y})$, was applied to the data above in 

$512$~sec long segments overlapped by $64$~sec. The last $64$~sec of the cross-correlation of a segment with a template are discarded to account for the corruption caused by the circular wrapping of the template in an FFT-based cross-correlation. Of the resulting $1135$ segments, three were dropped because they contained confirmed GW signals, namely, GW190706\_222641, GW190707\_093326, and GW191215\_223052. Assuming that our search method is not more sensitive than the flagship CBC search pipelines, this ensured that all candidate events detected by the PSO-based search in the remaining segments, in the absence 
of injected signals, were either glitches or
false alarms arising from the Gaussian noise component. 

For some of the segments, the candidate signal had a time of arrival $t_a$ in the part of the cross-correlation output that was not discarded but a time of coalescence, $t_a + \tau$ [c.f., Eq.~(\ref{eq:chirplengtheqn})], that fell in the discarded part. In most of these cases, this happened due to a strong glitch in the last $64$~sec of a segment triggering a high cross-correlation with a long-duration ($> 64$~sec) template.
Clearly, the estimated $t_a$, which is integral to \algoname, is invalid for such a candidate and the corresponding segment was dropped from further consideration. However,
this did not result in the loss of a glitch
because it was caught, along with valid estimated parameters,
in the overlapped section at the start 
of the subsequent segment. As a result of this cut, an additional $68$ segments were removed, reducing the duration of the analyzed data to $\approx 131$ hours. The dropping of the above segments was a straightforward solution given that the data chosen has no real GW signal. In a more sophisticated search implementation, one could simply 
repeat the search by shifting the segment boundaries 
such that both the estimated $t_a$ and $t_a+\tau$ of a candidate 
event are in the nonoverlapped part.

Prior to carrying out the PSO-based search, each data segment was high-pass filtered with a cutoff frequency of $30$~Hz to remove the seismic noise contribution.  
For every $512$~sec data segment, each run ($S_P$ and $S_N$) took $\approx 46$~min on $8$ cores of an AMD EPYC 7763 64-Core Processor running at $2.45$ GHz. 
\subsection{Signal injections}
\label{sec:siginj}
Each segment in the final set above
was first analyzed with a positive quadrant, $S_P$, search
without any signal injection and the candidate
event, with one event per segment, was recorded. Next, to study the veto safety of \algoname,
simulated GW signals were injected in a randomly chosen subset of segments in which
the estimated SNRs of the candidate event
fell below a threshold of $8.0$, thereby ensuring that these segments either
did not have glitches or had weak and inconsequential ones at most.
From the $321$ such segments that were found, we randomly picked, with replacement,
$454$ segments for signal injection. The set of injected signals
was comprised of low-, intermediate-, and high-mass systems, with masses defined in the detector frame, that were
distributed in SNR and other parameters as described below.

The low-mass injections had each mass chosen independently from a uniform distribution over $[1.4,3] \textup{M}_{\odot}$.
For intermediate-mass injections, the lower and higher masses were drawn from uniform distributions over $[1.4,3]$~$ \textup{M}_{\odot}$ and $[5,10]$~$ \textup{M}_{\odot}$, respectively. For the high mass injections, each mass was selected from a uniform mass distribution over $[10,25] \textup{M}_{\odot}$.
For each signal injection, the initial phase was uniformly sampled from a range of $[0,2 \pi]$ and the TOA was randomly selected to be between $[50,350]$ ~sec.
The TOA range ensured that the injected signal did not have a time of coalescence that
leaked into the overlap region between consecutive segments. 
The SNRs of each signal was drawn independently of the other parameters from uniform
distributions over three different ranges: 
  $198$ injections were drawn from $[10,40]$, which is representative of the SNR range of the GW signals that have been observed so far;  $196$ injections had a lower SNR range of $[10,13]$ and $60$ were injected with unrealistically high SNRs in $[100,500]$. The last set of injections was intended to stress test the assumption in \algoname that a GW signal will always show a large SNR contrast in searches with physical and unphysical waveforms. It is conceivable that this assumption is violated for very loud GW signals since they may induce a strong response from unphysical waveforms just as strong glitches can trigger a high-SNR response from physical waveforms. To illustrate that there is no loss of generality in restricting our injections to certain mass bands above, we inject $18$ signals with masses outside the bands described above. In the following, we refer to such signals as mass-gap signals.

   In addition to the main set of signal injections above, we tested the veto safety of \algoname with signals corresponding to larger mass components. In the first set of the additional injections, the masses were each chosen to be in $ [25, 40]$~$\textup{M}_\odot$ and the same SNR ranges as the main set were used. Although well outside the range of validity of the 2PN approximation, the generated signals still look like binary inspirals with a well-defined increasing instantaneous frequency and amplitude. This allows
  a qualitative assessment of the bounding values of masses for which the veto safety of \algoname remains intact. We injected a total of $140$ signals in
  this set.

 In the second set of additional injections, we used signals
 with waveforms belonging to the IMRPhenomXHM~\cite{PhysRevD.102.064002} family available in the PyCBC package~\cite{alex_nitz_2024_10473621} that models the binary inspiral signal more faithfully at higher masses. This allows us to do a test of veto safety that is independent of the one with 2PN signals and also extends the test to much larger component masses. However, an important caveat that should be kept in mind here is that the 2PN templates cannot achieve the highest possible match with IMRPhenomXHM waveforms, especially as the masses become larger. Nonetheless, this test provides a preliminary indication of how the performance of \algoname scales with component masses. The component masses for the IMRPhenomXHM injections include the complete mass range of the 2PN injections as well as masses in the range corresponding to the extreme high-mass systems that have been observed to date~\cite{PhysRevD.109.022001}, namely, $m_1 \in [40, 80]$~$\textup{M}_\odot$ and $m_2 \in [80, 110]$~$\textup{M}_\odot$. 
  For each mass range, including the extreme high-mass one, we injected $20$ signals in each of the three SNR ranges noted earlier. This results in $60$ injected signals for each mass range, resulting in a total of $300$ IMRPhenomXHM injections.


\subsection{Veto performance}
\label{sec:veto_results}
\begin{figure} 
    \centering
    \includegraphics[scale=0.16]{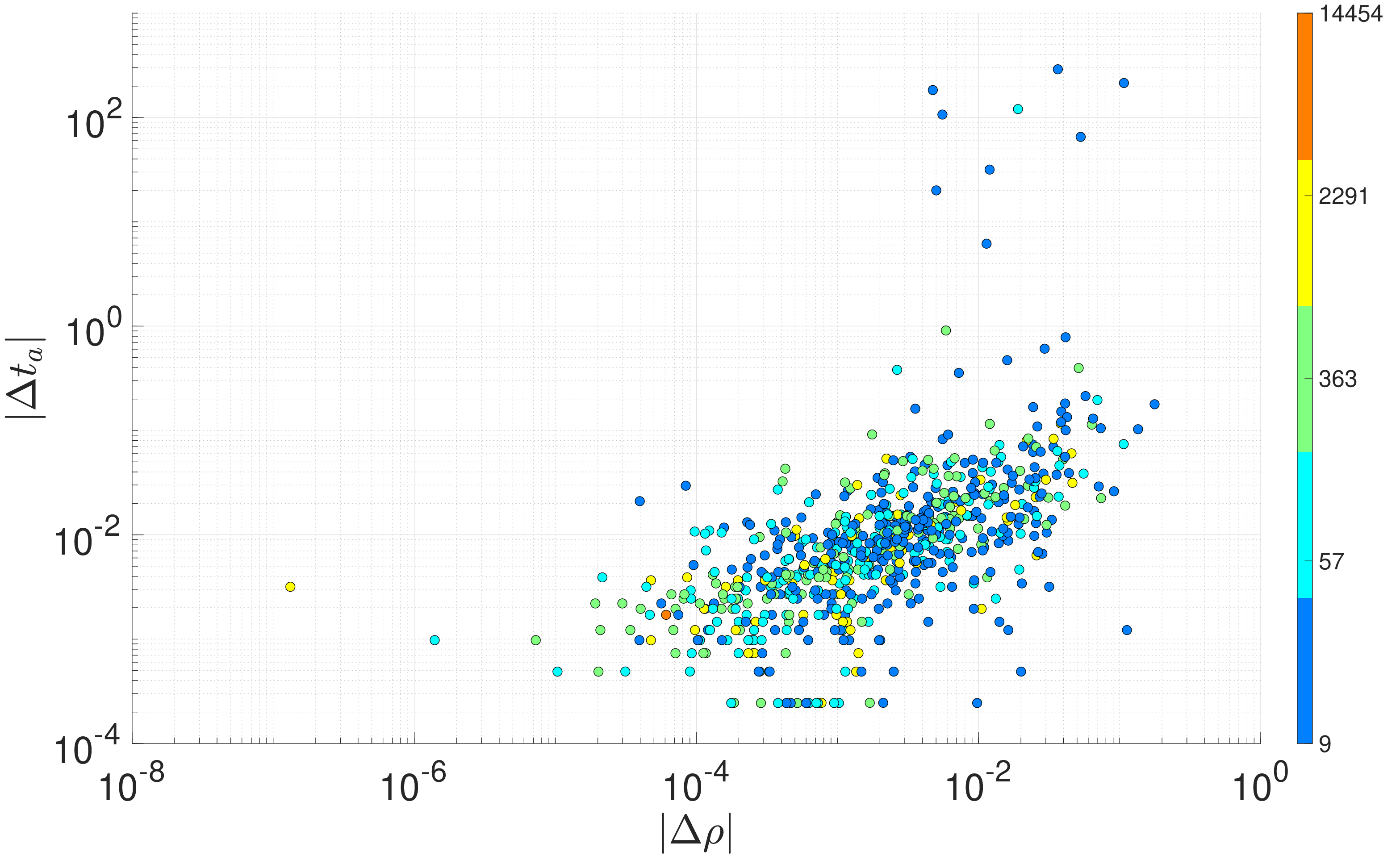}
    \caption{Scatterplot of glitches in the $(|\Delta \rho|,|\Delta t_a|)$ plane. The color of a point represents, following the map shown
    in the colorbar, the estimated SNR ($\widehat{\rho}_P$) in $S_P$ of a glitch. The colormap is based on the range of observed $\log(\widehat{\rho}_P)$ divided into $5$ bins.}
    \label{fig:alldata_distr}
\end{figure}

Candidate events were obtained from the data segments described in Sec.~\ref{sec:data_describe} using a 
detection threshold of $\widehat{\rho}_P\geq 9.0$ on the estimated SNR in the positive chirp time quadrant ($S_P$) search. (This threshold is independent of the one used to select the segments for signal injection as described in Sec.~\ref{sec:siginj}.)
 The probability of a false alarm from Gaussian stationary noise alone would be very low  ($< 1$~event/year) at such a high threshold~\cite{PhysRevD.54.7108} although real broadband noise (coupled with imperfect whitening) may have a non-Gaussian tail~\cite{PhysRevD.93.082005}
that elevates this probability somewhat. Hence, from the segments free of signal injections, most candidate events  are expected to be glitches and it is
 convenient to refer to them as such in the following. The objective of any veto method should be the  rejection of as many of these glitches as possible.
The set of candidate events found had $708$ glitches and $450$ injected 
signals from the main set. The four injections that were lost, corresponding to
a false dismissal probability of $0.8\%$, had injected SNRs in the 
lowest range ($[10,13]$) and estimated SNRs of $8.02$, $8.31$, $8.90$, and $7.86$. From 
the injections in the $m_1, m_2\in[25, 40]$~$\textup{M}_\odot$ set, $2$ injections were lost, both of which had low injected SNRs of $8.02$ and $8.56$. While \algoname is applied in stages as described in Sec.~\ref{sec:veto_strategy}, it is informative to apply the negative chirp time quadrant ($S_N$) search to all the glitches and obtain their $|\Delta t_a|$ [Eq.~(\ref{eq:delta_ta_def})] and
$|\Delta \rho|$ [Eq.~(\ref{eq:delta_rho_def})] values. 
Figure \ref{fig:alldata_distr} shows the distribution of all the glitches in the $(|\Delta \rho|, |\Delta t_a|)$ plane along with the distribution of their $S_P$ estimated SNRs. In agreement with the intuitive motivation for choosing $|\Delta t_a|$ and
$|\Delta \rho|$ presented in Sec.~\ref{sec:veto_strategy},  most of the glitches cluster
in the region with small values of these quantities. However,  a few distinct outliers are observed in the
region defined by $|\Delta t_a|\gtrsim 1$. We also see from Fig.~\ref{fig:alldata_distr} that
there is no notable dependence on SNR of the joint distribution in
$|\Delta t_a|$ and $|\Delta \rho|$ in the main cluster of glitches. On the other hand, most of the outliers have low SNRs.

\begin{figure}
    \centering
    \includegraphics[scale = 0.20]{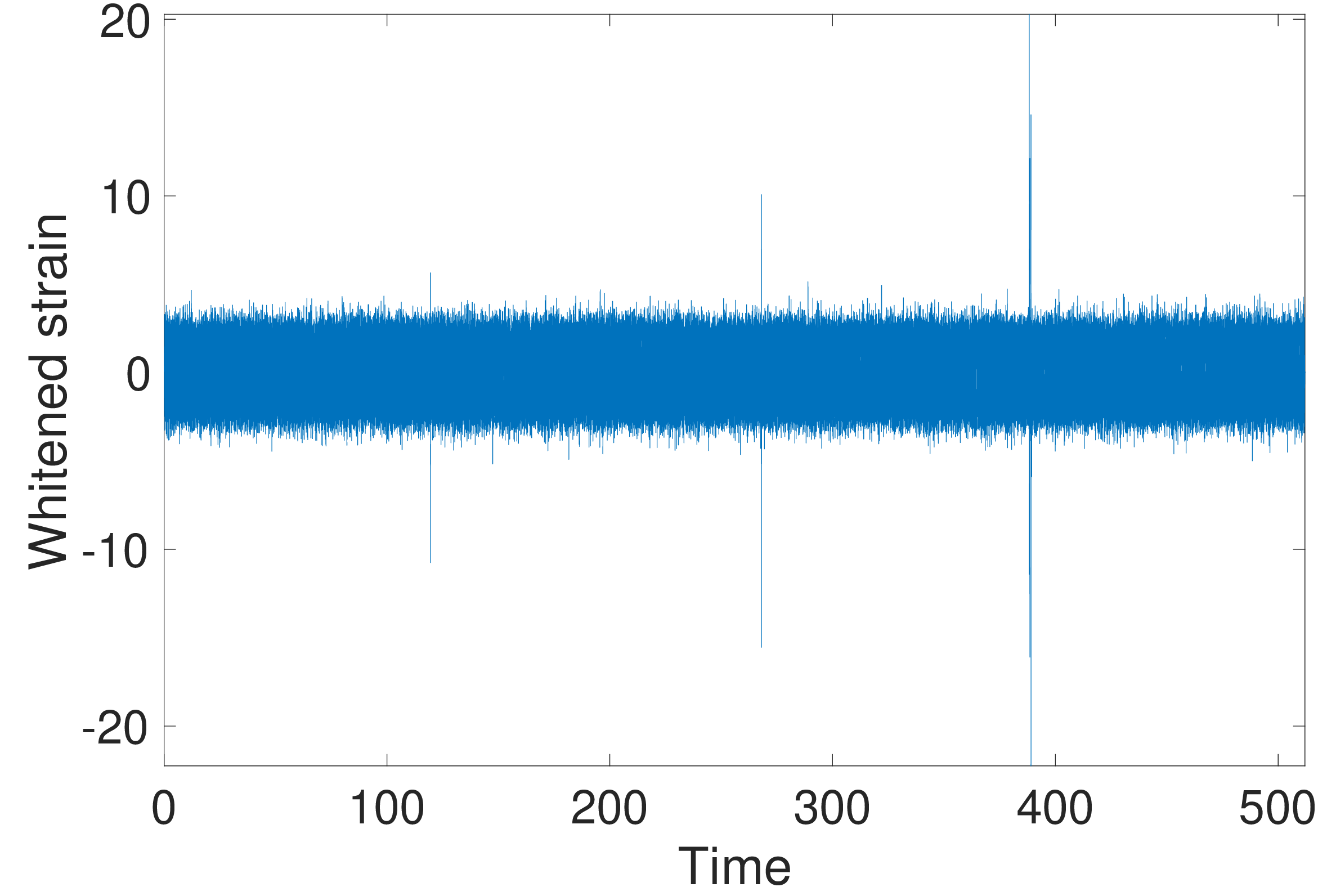}
    \caption{Whitened time-series of a data segment containing multiple glitches (high-amplitude spikes). The $S_P$ search detects the glitch located near $388$~sec, while the $S_N$ search detects the glitch near $268$~sec. The estimated SNRs of both glitches were $\approx 66$ in the respective searches.
    }
    \label{fig:specialcase}
\end{figure}

 The origins of these outliers, deciphered by closer inspection of the data segments that contain them, were found to be of $4$
different types. 
The origins of these outliers, deciphered by closer inspection of the data segments that contain them, were found to be of four
different types. 
\renewcommand{\theenumi}{\roman{enumi}}%
\begin{enumerate}
\item{\gtype{1}--} Multiple widely separated strong glitches in the same data segment cause the $S_P$ and $S_N$ searches to detect different ones. An example is shown in Fig.~\ref{fig:specialcase}. 
\item{ \gtype{2}--} The $S_P$ and $S_N$ searches find events associated with a single
but wide duration ($\mathcal{O}(10)$~sec) disturbance that causes them to estimate similar SNRs but widely different TOAs.
\item{ \gtype{3}--} The data segment has no obvious glitch, but creates a
false alarm in the $S_P$ search, possibly due to broadband non-Gaussianity mentioned earlier. Since there is no glitch, only broadband noise, the $S_N$ search finds an event that is widely separated in TOA. 
\item{\gtype{4}--} The $S_P$ search detects a weak glitch but the $S_N$ search does not and finds an event associated with broadband noise. This causes a large discrepancy in both SNR and TOA.
\end{enumerate}
In the set of outlier glitches, there were $4$, $2$, $2$, and $1$ \gtype{1} to \gtype{4}
events, respectively. \gtype{1} and \gtype{2} events arise from a technical limitation of the current PSO-based search since it only finds the global maximum of the fitness function and, hence, identifies only one candidate per data segment in a $S_P$ or $S_N$ search. Strictly speaking, \gtype{3}
glitches are simply false alarms whose prevalence depends on the detection threshold used for the $S_P$ search and the non-Gaussianity of the broadband noise distribution.
Future improvements to the PSO-based search that allow multiple candidate events per segment and better accounting of broadband non-Gaussianity could mitigate these
types of glitches better. 
In the set of outlier glitches, there were $4$, $2$, $2$, and $1$ \gtype{1} to \gtype{4}
events, respectively. \gtype{1} and \gtype{2} events arise from a technical limitation of the current PSO-based search since it only finds the global maximum of the fitness function and, hence, identifies only one candidate per data segment in a $S_P$ or $S_N$ search. Strictly speaking, \gtype{3}
glitches are simply false alarms whose prevalence depends on the detection threshold used for the $S_P$ search and the non-Gaussianity of the broadband noise distribution.
Future improvements to the PSO-based search that allow multiple candidate events per segment and better accounting of broadband non-Gaussianity could mitigate these
types of glitches better. 

\begin{figure*}
    \centering
    \includegraphics[scale = 0.3]{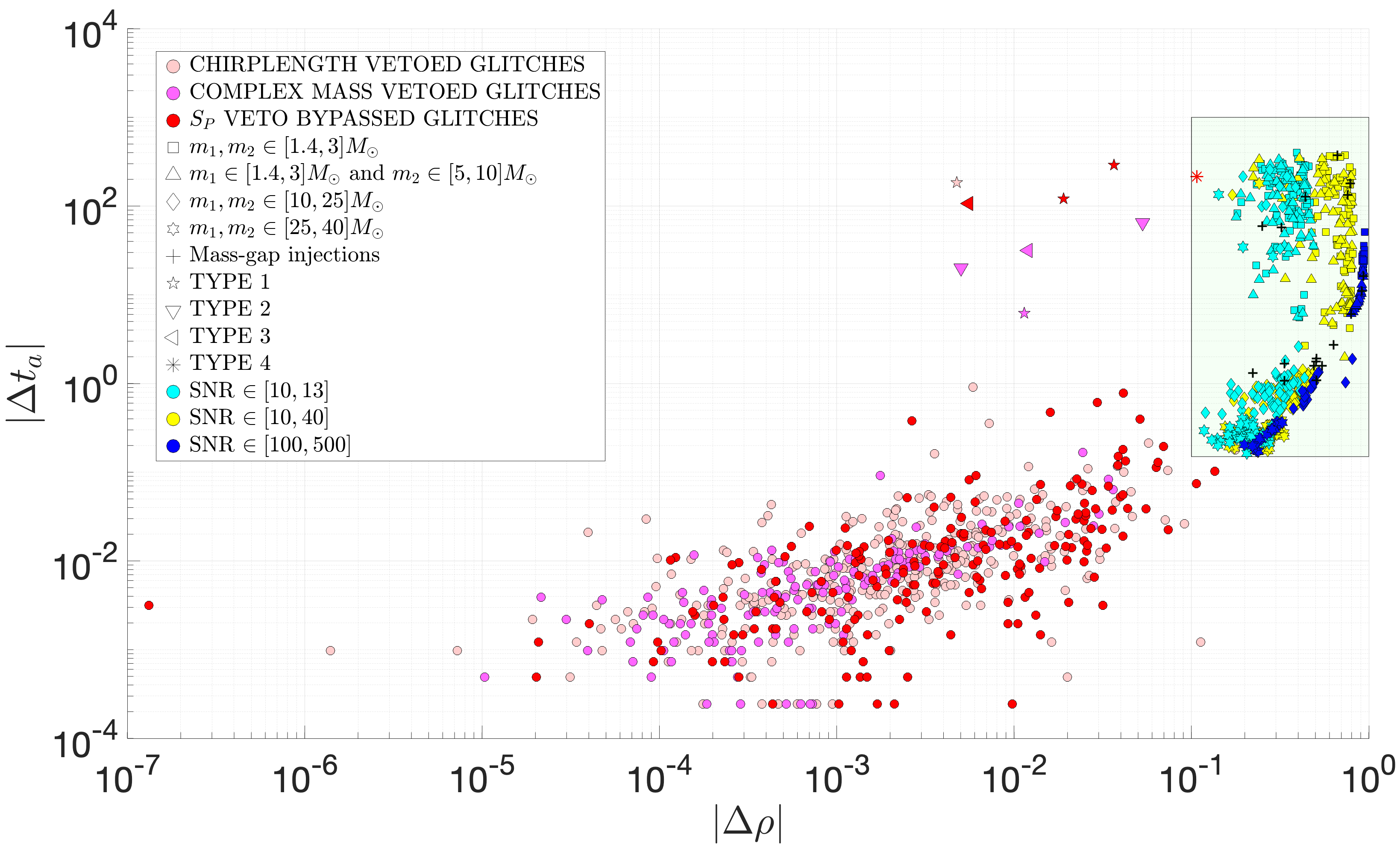}
    \caption{ Distribution in the $(|\Delta \rho|,|\Delta t_a|)$ plane of candidate events obtained with a detection threshold of ${\rm SNR}=9.0$. 
    Glitches vetoed by the chirp length and complex mass veto steps of \algoname in the positive chirp time quadrant ($S_P$) are shown with pink and magenta colors, respectively.  
    The remaining glitch events that bypass the $S_P$ vetoes are shown in red. The exclusion area for the $S_N$ veto step, defined by $|\Delta \rho| \geq 0.1$ and 
    $|\Delta t_a| \geq 0.15$ is shown as a rectangle shaded in green. Different marker shapes for the glitches correspond to different types of outliers, as noted in the legend of the plot. Candidate events from injected signals are shown in dark blue, yellow, and cyan colors, corresponding to the high, realistic, and low SNR ranges, respectively. Different marker shapes for these events correspond to different mass ranges, as noted in the legend. The additional detected signal injections for the mass gap ranges of $m_1,m_2 \in [3,10] \textup{M}_{\odot}$ and $m_1 \in [1.4,3] \textup{M}_{\odot}$, $m_2 \in [10,25] \textup{M}_{\odot}$ are plotted as black `+' markers, whereas the detected injections in the range of $m_1,m_2 \in [25,40] \textup{M}_{\odot}$ are plotted as hexagrams. The single glitch that falls into the exclusion zone and is not vetoed by the $S_P$ vetoes is a \gtype{4} outlier shown by the red asterisk.}
    \label{fig:posspace_vetoedpoints}
\end{figure*}

\begin{figure}
    \centering
    \includegraphics[scale = 0.15]{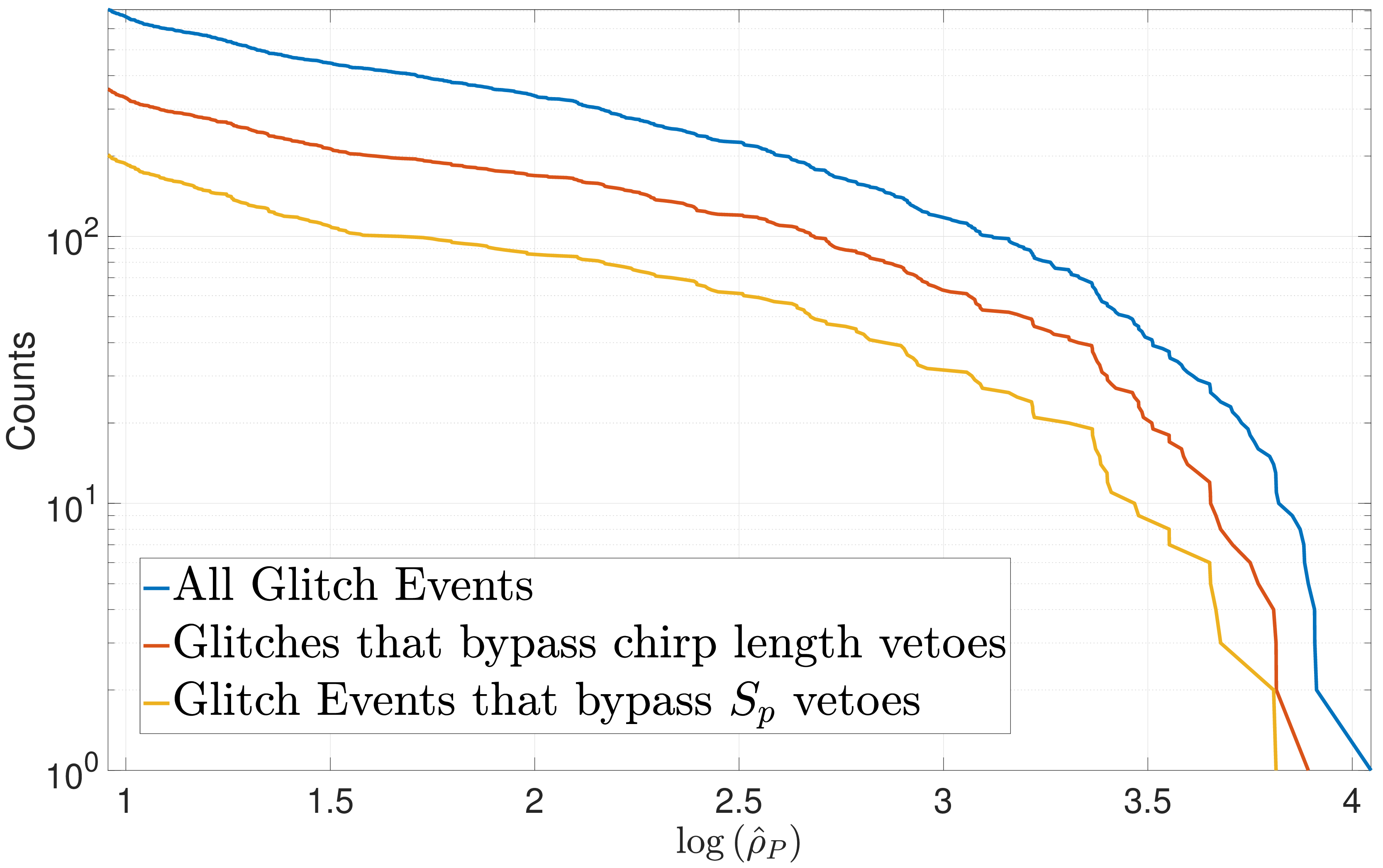}
    \caption{The survival function for  estimated SNRs in the $S_P$ search, $\widehat{\rho}_{P}$, of all glitches (blue), glitches that bypass the chirp length veto step (orange), and glitches that bypass all positive chirp time 
    quadrant ($S_P$) vetoes (yellow). For any given value of $\widehat{\rho}_P$, the survival function provides the number of glitches with estimated SNRs above this value.}
    \label{fig:histogram}
\end{figure}

Figure~\ref{fig:posspace_vetoedpoints} presents the main results on the performance of
\algoname. They are conveniently summarized again in terms of a scatterplot in the $|\Delta \rho|$, $|\Delta t_a|$ plane, with the detected injections now included in addition to glitches. The points have also been differentiated on the basis of the different veto steps in \algoname, with glitches vetoed by the three steps described in Sec.~\ref{sec:veto_strategy} shown with different colors. We have also shown the different types of outliers.  With the two $S_P$ vetoes alone, namely chirp length and complex mass,  $505$  out of the $708$ glitches, or $71.4\%$ of the total number of glitches,  were vetoed with $0 \%$ rejection of injected signals. Out of these $505$ glitches, $351$, or $49.6\%$ of the total number of glitches, were vetoed by the chirp length veto alone. These two vetoes also rejected five out of the nine outliers discussed above. Figure~\ref{fig:histogram} shows the fraction of glitches that survive after the application of the chirp length and complex mass vetoes as a function of their estimated SNR $\widehat{\rho}_P$. We see that each of the two $S_P$ vetoes had nearly a constant effectiveness, in terms of the fraction of vetoed glitches,  across nearly the entire range of glitch SNRs. 

\begin{figure}
    \centering
    \includegraphics[scale = 0.16]{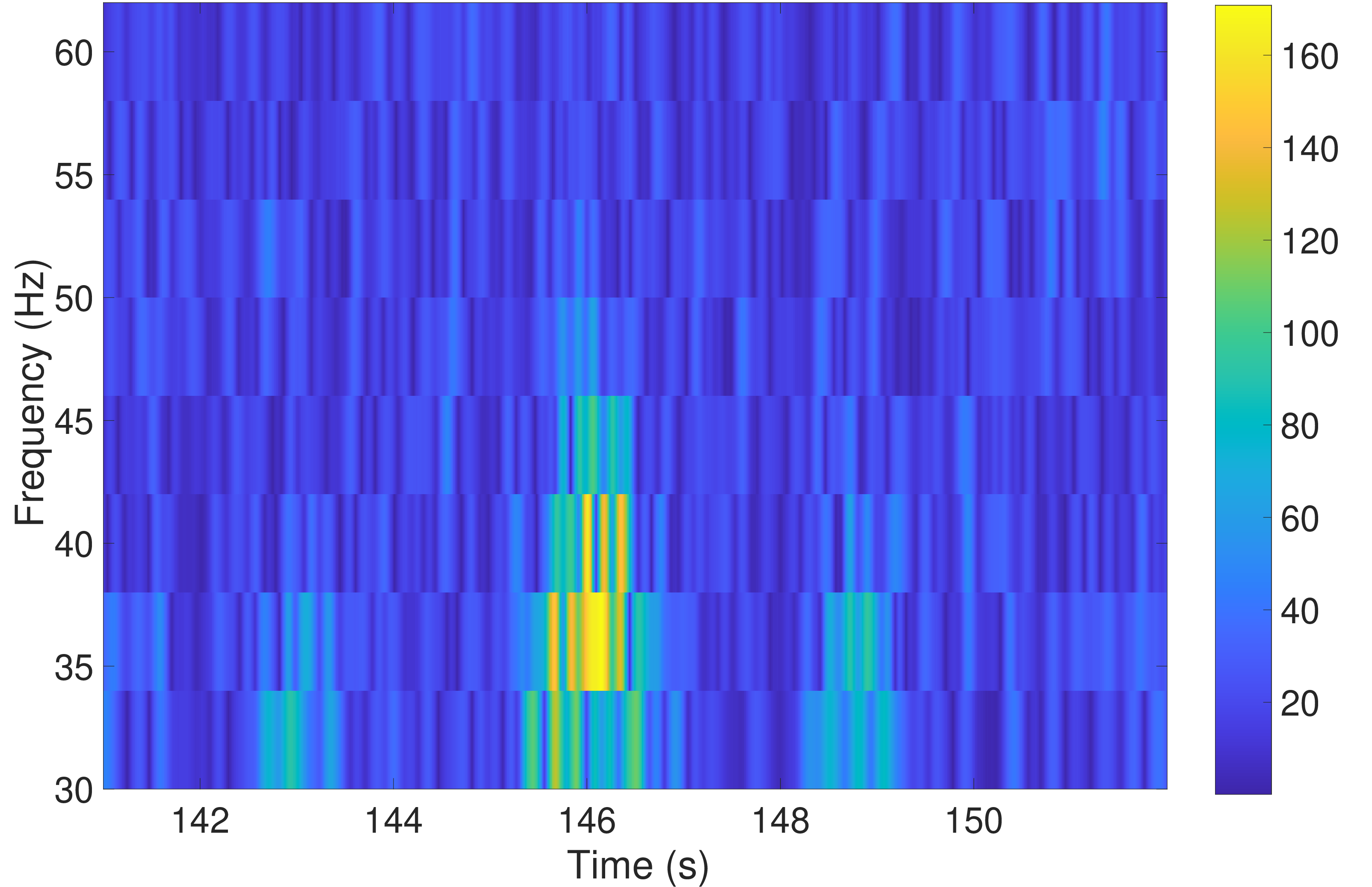}
    \caption{Spectrogram of a short segment of whitened data containing the glitch, a \gtype{4} outlier, that survived \algoname. The spectrogram is restricted in the frequency range of $[30,60]$ Hz and is plotted on a linear scale. The positive chirp time quadrant ($S_P$) search detects the low-frequency low-SNR glitch near $145$ ~sec while the negative chirp time quadrant ($S_N$) search produces only a broadband noise-induced event that is widely separated in time. 
    }
    \label{fig:glitchanomalyspectrogram}
\end{figure}

For the third veto step in \algoname based on the $S_N$ search,  the observed distribution of detected injections 
in the $(|\Delta \rho|, |\Delta t_a| )$ plane
can be used to define an exclusion area
such that any candidate event falling outside it is vetoed. 
In the present case, letting this area enclose all detected injections up to component masses of $40$~$\textup{M}_\odot$ defines its boundary as
$|\Delta \rho| \geq 0.1$ and $|\Delta t_a| \geq 0.15$. 
As seen in Fig.~\ref{fig:posspace_vetoedpoints}, imposing this criterion rejects all
remaining glitches except for the single \gtype{4} outlier. This event was a marginal detection in the $S_P$ search with an estimated ${\rm SNR}=9.12$ and, as shown in  Fig.~\ref{fig:glitchanomalyspectrogram}, arises from a visible low-frequency glitch at the edge of the high-pass filter cutoff. It bypassed both the chirp length and complex mass vetoes in $S_P$, with estimated masses of $4.33 \textup{M}_{\odot}$ and $32.75 \textup{M}_{\odot}$.
Such glitches, although rare,  may be an important contaminant in CBC searches, and it would be worth characterizing their population with much larger datasets. Figure \ref{fig:posspace_vetoedpoints} also
shows mass-gap injections, with their locations in the $(|\Delta \rho|,|\Delta t_a|$) plane falling 
in between those of the main set of injections in the low-, intermediate-, and high-mass bands. This 
shows that there is no loss in the generalization
of the above results to masses outside the mass bands.

\begin{figure}
    \centering
    \includegraphics[scale = 0.16]{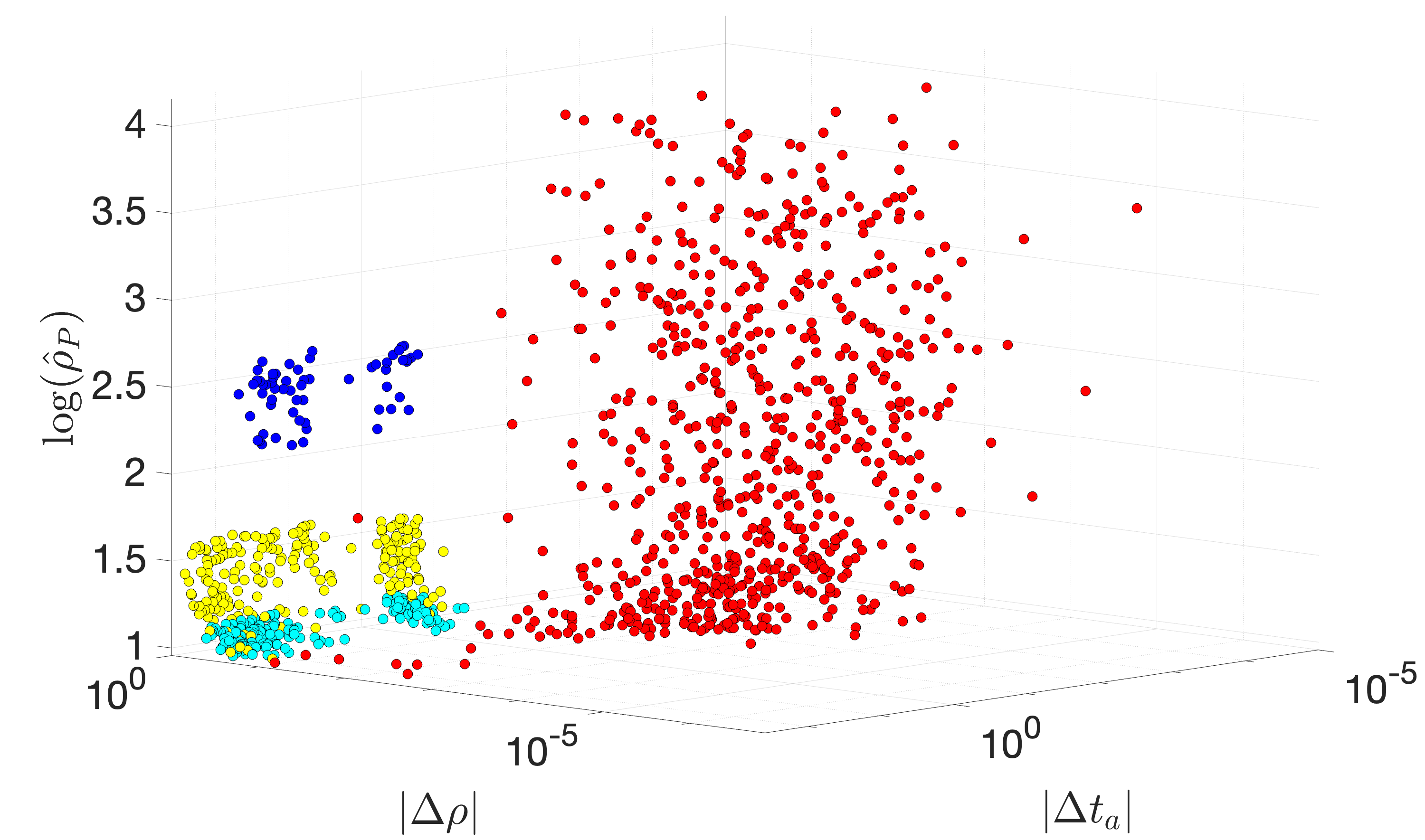}
    \caption{Distribution of candidate events in the 3-dimensional $(|\Delta \rho|,|\Delta t_a|,\log{(\widehat{\rho}_P)})$ space. Red dots show glitches while injected signals are shown in the same color scheme as used in Fig.~\ref{fig:posspace_vetoedpoints}.}
    \label{fig:3dplot}
\end{figure}

\begin{figure}
    \centering
    \includegraphics[scale = 0.15]{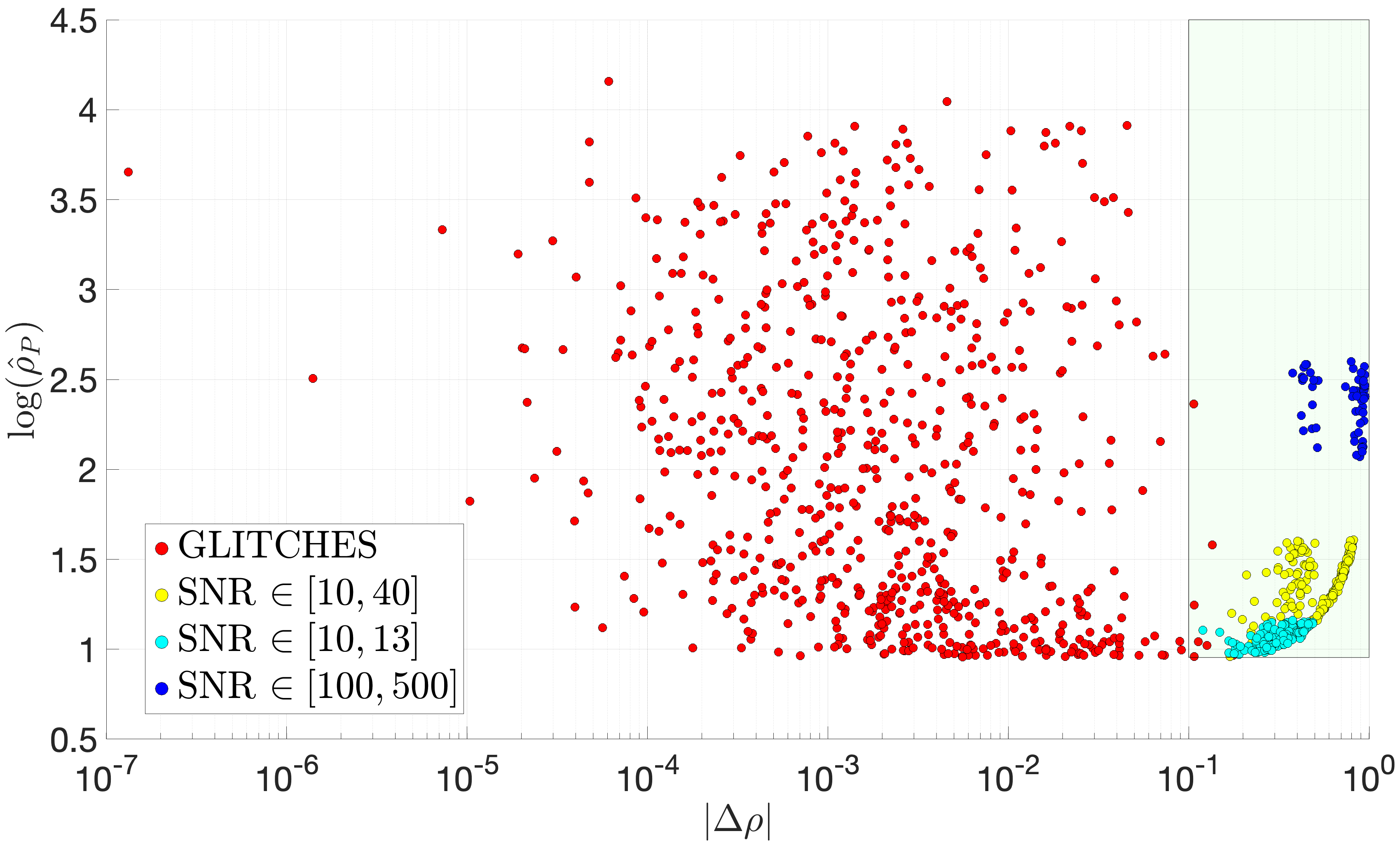}
    \caption{Distribution of candidate events in $(|\Delta \rho|,\log{(\widehat{\rho}_{P})})$ plane. Red dots show glitches while injected signals are shown in the same color scheme as used in Fig.~\ref{fig:posspace_vetoedpoints}. The shaded region is bounded from below by the detection threshold and from left by the threshold on 
    $|\Delta \rho|$. }
    \label{fig:lognsrvdeltasnr}
\end{figure}

 In addition to  the  two-dimensional distribution of candidate events in the
 $(|\Delta \rho|,|\Delta t_a|)$ plane, it is illuminating to examine their three-dimensional distribution, shown in Fig.~\ref{fig:3dplot}, 
 with the third dimension being $\log{(\widehat{\rho}_{P})}$. (For visual clarity, the injections in the $[25, 40]$~$\textup{M}_\odot$ range and the mass-gap injections are not included in this and subsequent figures.)  The segregation of the injections in SNR is more apparent whereas they overlap in the two-dimensional plot.
 The glitches are well separated from the injections and form a diffuse and 
 unstructured cloud of points. On the other hand, the cloud of injection
 points is highly structured. This is revealed more clearly in Fig.~\ref{fig:lognsrvdeltasnr} where the three-dimensional distribution is projected onto the $(|\Delta\rho|, \log(\widehat{\rho}_P))$ plane: we see that the injections lie in the neighborhood of a two-dimensional surface embedded in the three-dimensional space. The elucidation of the origin of this structured distribution is left to future work.

With $99.9\%$ of glitches vetoed successfully and $0 \%$ loss of detected injections, our results shows that \algoname is highly effective and has a  high level of veto safety up to a total mass of $80$~$\textup{M}_\odot$.
In terms of the inverse false alarm rate (IFAR) metric used for detection 
confidence in GW discoveries, the IFAR was reduced by the application of \algoname from $1$~false alarm every $11.10$~min to $1$~false alarm every $5.45$~days.
  The breakdown of the 
 veto fraction across the three veto steps shows that the $S_N$ veto step in \algoname is essential as it is required to eliminate $\approx 30\%$ of glitches. At the same time, this shows that it is not needed for the majority of candidate events, thereby considerably reducing its impact on the overall computational cost of the search.
The tunable parameters in \algoname are (i) the threshold on  
$\zeta$  (Eq.~\ref{eq:imagratioexpression}) in the complex mass veto step, (ii)
ranges of $\tau_0$
and $\tau_{1.5}$ defining the $S_N$ space, and (iii) the thresholds on $|\Delta \rho|$
and $|\Delta t_a|$ in the $S_N$ veto step. (The chirp length veto step has no free parameters.)
For a PSO-based search, it is straightforward to 
set the ranges of $\tau_0$
and $\tau_{1.5}$ in $S_N$ to be the same as those for $S_P$ since there is no impact
of smaller ranges on the computational cost of the $S_N$ search. The remaining parameters can be fixed using the distribution of injections alone. For
 example, we set the thresholds on $|\Delta \rho|$ and $|\Delta t_a|$
 above such that there was $0\%$ loss of detected injections. Similarly, as shown in Table~\ref{tab:complexratios},
we examined the fraction of detected injections in the main set
vetoed by different thresholds on $\zeta$ and determined that it should be set 
at $90\%$ in order to retain all injections. Table~\ref{tab:complexratios}
also shows the fraction
of glitches that are missed by the complex mass veto for different thresholds
on $\zeta$. One can imagine a scenario in which
the threshold is lowered to tolerate a higher fraction of rejected injections while reducing computational costs due to a less frequent application of the $S_N$ search.  
 \begin{table}[t]
\label{tab:ratios}
    \centering
    \begin{tabular}{|c|S |S|}
    
    \hline
    \hline
     $\zeta$  &  {Glitches missed} &  {Injections rejected} \\
     ($\%$) & {($\%$)} & {($\%$)} \\
    \hline
     10 & 18.7 & 15.1 \\
     25 & 19.7 & 4.4 \\
     50 & 22.1 & 0.4 \\
     90 & 28.6 & 0.0 \\
     100 & 31.9 & 0.0 \\
    \hline
    \hline
\end{tabular}
    \caption{Effect of the threshold on $\zeta$ [Eq.~(\ref{eq:imagratioexpression})] on the fraction of glitches (second column) not rejected  and the fraction of detected injections (third column) rejected by the complex mass veto. 
    }
    \label{tab:complexratios}
\end{table}

\begin{figure*}
    \centering
    \includegraphics[scale = 0.30]{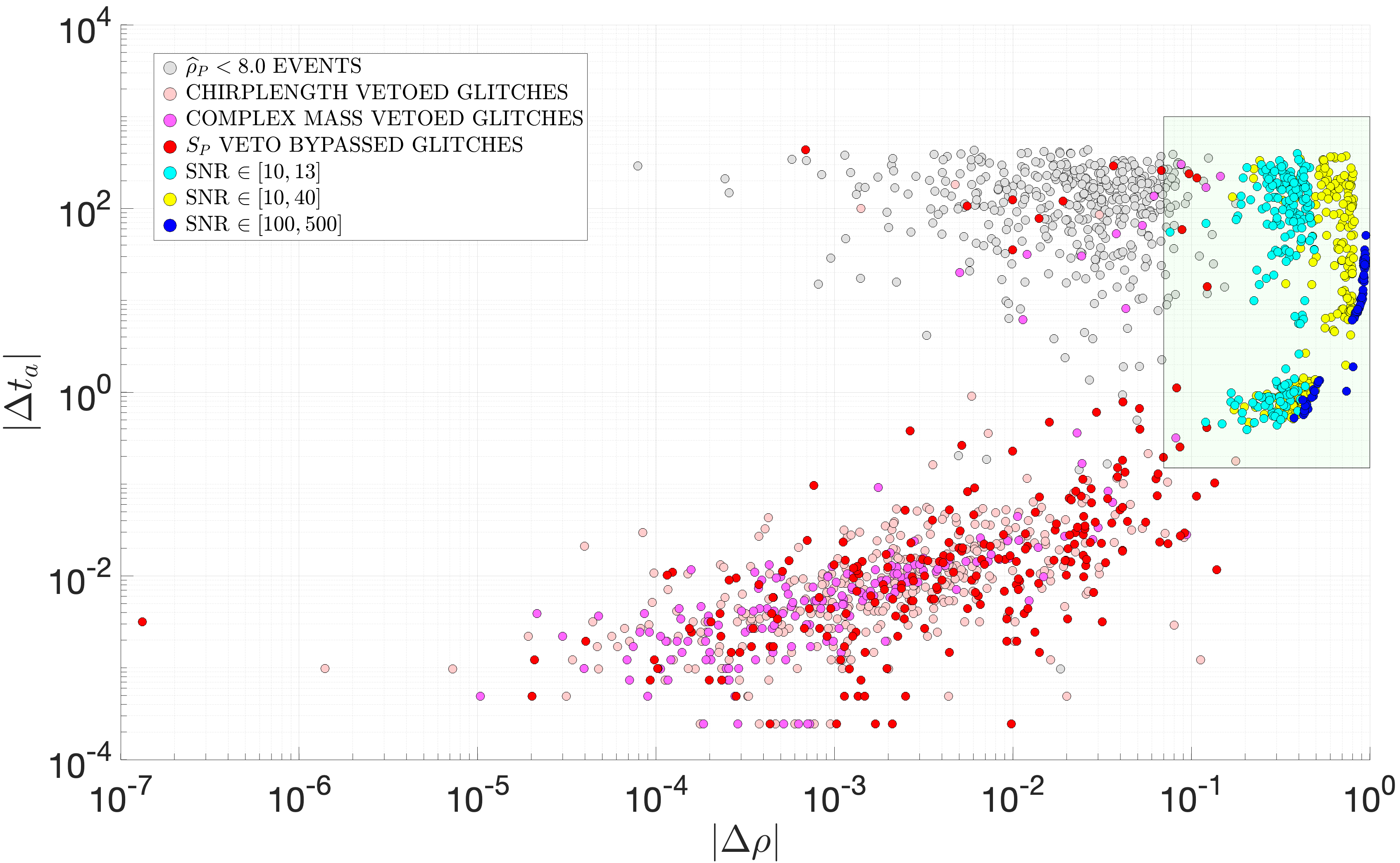}
    \caption{ Distribution in the $(|\Delta \rho|,|\Delta t_a|)$ plane of candidate events obtained with a detection threshold of ${\rm SNR}=8.0$. 
    Glitches vetoed by the chirp length and complex mass veto steps of \algoname in the positive chirp time quadrant ($S_P$) are shown with pink and magenta colors, respectively.  
    The remaining glitch events that bypass the $S_P$ vetoes are shown in red. The exclusion area for the $S_N$ veto step, defined by $|\Delta \rho| \geq 0.07$ and $|\Delta t_a| \geq 0.15$ is shown as the green shaded rectangle.  Candidate events from injected signals are shown in dark blue, yellow, and cyan colors, corresponding to the high, realistic, and low SNR ranges, respectively. Events 
    that fell below the detection threshold are shown in gray color.
    }
    \label{fig:detectionthreshold_8_vetoedpoints}
\end{figure*}

With an effective glitch veto strategy in hand, the detection threshold itself can be lowered to catch weaker signals. Fig.~\ref{fig:detectionthreshold_8_vetoedpoints} shows the results when the detection threshold is lowered to $\widehat{\rho}_P\geq8.0$. As intended, 
this allows the detection of three out of 
the four injected signals that had been missed with the higher threshold.
The number of crossings of the detection threshold now increase to $762$ candidate events from the segments without signal injections. (One expects that there will be
more contamination of this set by false alarms arising from the broadband noise, but we will continue to call them all glitches.)
\algoname was applied as before, with all settings kept the same except for the exclusion area in the $(|\Delta \rho|,|\Delta t_a|)$ plane, which must be enlarged to $|\Delta \rho| \geq 0.07$ and $|\Delta t_a| \geq 0.15$ in 
order to retain all the detected injections. Now, the number of glitches that are 
missed rises to $7$ from $1$, reducing the fraction of vetoed glitches to $99.2\%$.
The loss of detected injections remains at $0\%$ after the application of $\algoname$, maintaining its high safety. While the IFAR of the weakest detected injection increases in correspondence with the larger number of glitches, the reduction in IFAR due to \algoname is still quite large: from $1$~false alarm every $10.31$~min before vetoing to $1$~false alarm every $18.7$~hours after. Figure~\ref{fig:detectionthreshold_8_vetoedpoints} also shows the events that did not cross the detection threshold. Most of such events arise from the broadband noise rather than glitches, like the \gtype{3} outlier discussed earlier. We see that such noise-induced events occupy the same region of the $(|\Delta \rho|,|\Delta t_a|)$ plane as the outlier glitches and that this is well separated from the main cluster of  glitches.
Lowering the detection threshold further and expanding the exclusion zone to retain
weaker injected signals will uncover more of this population of
noise-induced candidate events, which are more likely to bypass
\algoname and, for that matter, any other glitch veto method.

One way in which \algoname differs from $\chi^2$-vetoes is that the 
latter depend primarily on the norm of the data projection on the null space of a given template, while \algoname uses both the projection norms, encapsulated in $|\Delta \rho|$, as well as the 
estimated TOAs, through 
 $|\Delta t_a|$, of the best fit physical and unphysical templates.
Figure~\ref{fig:posspace_vetoedpoints} and Fig.~\ref{fig:detectionthreshold_8_vetoedpoints} illustrate the 
importance of using $|\Delta t_a|$ in addition to $|\Delta \rho|$ in \algoname. 
If the threshold on 
$|\Delta t_a|$ is removed, one can see that the number of glitches 
missed (red colored points) increase by $2$ and $7$ in Fig.~\ref{fig:posspace_vetoedpoints} and Fig.~\ref{fig:detectionthreshold_8_vetoedpoints}, respectively, thereby 
increasing the false alarm rate due to glitches without a concomitant increase
in the number of detected injections. Another clear impact of using $|\Delta t_a|$, seen in Fig.~\ref{fig:detectionthreshold_8_vetoedpoints},
is the separation of glitches from below-threshold events (gray dots), the majority of
which would be false alarms from broadband noise and not real glitches. The separation includes both strong and weak (but above threshold) glitches since, as discussed in the
context of Fig.~\ref{fig:alldata_distr}, their distribution in the $(|\Delta \rho|,|\Delta t_a|)$ plane does not depend strongly on their estimated SNR. As the SNR of GW signals is reduced, they will migrate in this plane to the left, as can be seen by the progression of the colored bands of the injected signals, but not towards the bottom where most of the glitches reside.

\begin{figure*}
    \centering
    \includegraphics[scale = 0.30]{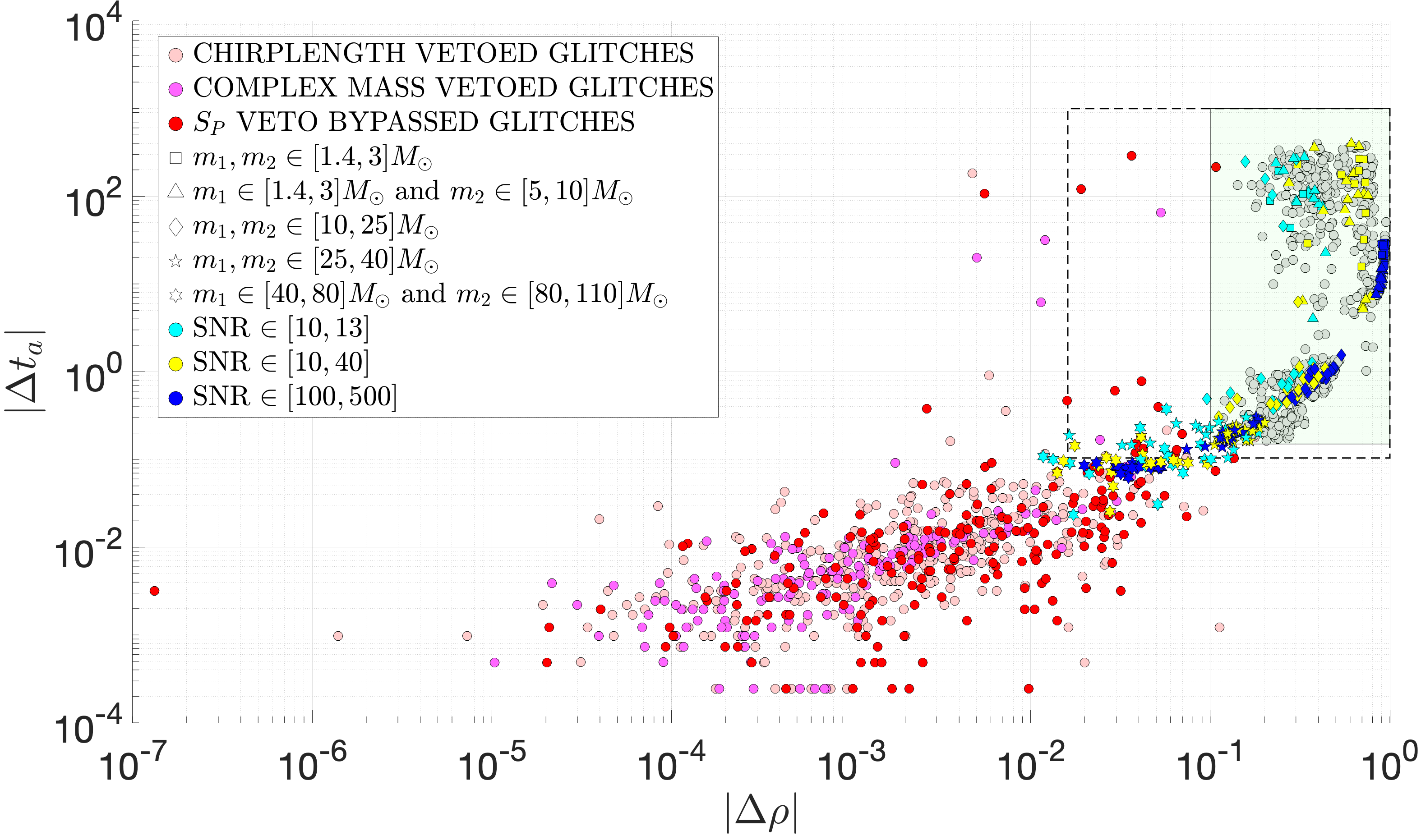}
    \caption{ Distribution in the $(|\Delta \rho|,|\Delta t_a|)$ plane of candidate events using  IMRPhenomXHM injected signals  and a detection threshold of ${\rm SNR} = 9.0$. The detected 2PN signal injections in Fig.~\ref{fig:posspace_vetoedpoints} are plotted here in light gray. Here, the glitches are differentiated using the same color code as in Fig.~\ref{fig:posspace_vetoedpoints} based on the type of $S_P$ and $S_N$ veto steps used for their rejection. Candidate events from injected IMRPhenomXHM signals are shown in dark blue, yellow, and cyan colors, corresponding to high, realistic, and low SNR ranges, respectively. Different marker shapes for these events correspond to different mass ranges, as noted in the legend. The green shaded rectangle is the same as in Fig.~\ref{fig:posspace_vetoedpoints}. The  rectangle with the dashed border, enclosing a region of $|\Delta \rho| \geq 0.0161$ and $|\Delta t_a| \geq 0.104$, includes all the leaked signal injections in the mass range  $m_1, m_2 \in [25,40] \textup{M}_{\odot}$.
    }
    \label{fig:vhminj_distrplot}
\end{figure*}

Finally, Fig.~\ref{fig:vhminj_distrplot} shows the distribution of the IMRPhenomXHM signal injections in the $( |\Delta \rho|, |\Delta t_a|)$ plane that were detected above the threshold of ${\rm SNR} = 9.0$. We see that most of the signals corresponding to a total mass of $\leq 80$~$\textup{M}_\odot$ occupy the 
same region as the 2PN signals. In the set of injections having the mass range $m_1,m_2 \in [25,40] \textup{M}_{\odot}$, $11$ injections in the lowest SNR range of $[10,13]$  leak out of the $S_N$ veto exclusion area. This is likely due to the larger mismatch between the 2PN templates and the IMRPhenomXHM signals at higher masses. In order to enclose these leaked injections, the boundary of the exclusion area must be expanded from $|\Delta \rho| \geq 0.1$ and $|\Delta t_a| \geq 0.15$ to $|\Delta \rho| \geq 0.0161$ and $|\Delta t_a| \geq 0.104$. Upon this expansion, an additional $11$ glitches are missed by \algoname, increasing the total number of glitches missed from $1$ to $12$. Therefore, the number of vetoed glitches falls from $707$ to $696$, decreasing the fraction of successfully vetoed glitches from $99.9\%$ to $98.3\%$. 
Going to the extreme high-mass systems, we see that the majority of the injections leak out of the expanded exclusion area. The bulk of these leaked injections lie in  $0.0115 \leq |\Delta \rho| \leq 0.0161$ and $0.0195 \leq |\Delta t_a| \leq 0.104$ although the actual distribution is better enclosed by a smaller nonrectangular area. A rough count showed that $\approx 28$ additional glitches, unvetoed by the $S_P$ veto steps, overlap with the leaked injections. Thus, preventing these injections from being vetoed would result in the number of vetoed glitches to fall from $696$ to $\approx 668$, reducing the fraction of successfully vetoed glitches to $\approx 94.3\%$ and increasing the IFAR from $1$ false alarm every $5.45$ days to $1$ every $\approx 3.3$ hours for the data used here. Alternatively, all of the leaked signals would be rejected by \algoname as glitches if the exclusion zone is not changed. 
It should be emphasized that this does not undermine the detectability of such signals because they will be detected in parallel searches using more refined template families. As discussed earlier, the \algoname approach of finding and using unphysical regions of the signal parameter space
needs to be extended to these 
template families in order to reliably assess its safety for high-mass signals.


\section{Conclusions}
\label{sec:conclusions}
We have presented a veto method called \algoname that exploits the unphysical regions of the positive chirp time quadrant, $S_P$, and unphysical waveforms in the negative chirp time quadrant, $S_N$, to separate glitches from CBC signals in a 
matched filter search. We have tested the effectiveness of our strategy using $\approx 131$~hours of single-detector LIGO data. Our results show that \algoname
is a highly effective veto method that can successfully reject glitches that impact CBC searches  while maintaining a high level of veto safety up to a total mass of $\approx 80$~$\textup{M}_\odot$. The search over the unphysical 
sectors of $S_P$ is already integrated in a PSO-based approach and does not add any 
computational burden. The search over $S_N$ incurs additional computational cost but need only be used for $\approx 30\%$ of candidate events. 
For the detection threshold of ${\rm SNR}=9.0$, \algoname eliminated all glitches except one and
brought down the IFAR due to glitches by a factor of $\approx 707$. It should be emphasized here that this large reduction was obtained from just
a single-detector search.
Supplementing with a multi-detector coincidence veto will reduce the IFAR even further. The gain from using \algoname in such a scenario will be quantified 
in future work.

In this study, a common set of injected signals was 
used for tuning the parameters of \algoname.
However, the injections needed to tune the complex mass veto threshold $\zeta$ should be more specialized. This is because the problem of complex estimated masses mainly affects the signals located on the boundary of the unphysical region of complex masses, 
which is given by the curve of equal mass binaries (cf. Fig.~\ref{fig:postauspace}). Therefore, a more precise
determination of $\zeta$ for a given fractional loss
of injected signals requires a much
larger number of equal-mass injections than were present in the common injection set.
As a result, we used a conservative value for $\zeta$ that incurred a larger fraction of missed glitches in the complex mass veto step. While future applications of \algoname will improve in this aspect, our current results are not significantly affected given that the $S_N$ veto step would have trapped any glitch that escaped the complex mass veto due to a lower $\zeta$. 

From  the outliers in the distribution of glitches across the $(|\Delta \rho|,|\Delta t_a|$) plane, we found that the majority are 
associated with a technical limitation of the current PSO-based search that allows
only one event to be identified per data segment. This limitation will be lifted in ongoing improvements
to the PSO-based CBC search that will enable the identification of 
all local maxima of the fitness function above a given detection threshold 
rather than just the global maximum. To aid reproducibility of our work, we have provided the dataset of candidate events found in our study, the codes for PSO-based matched-filtering, and the scripts for producing some of the main plots in a public data release on Zenodo~\cite{girgaonkar_2024_10547476}.

\acknowledgments
We thank the anonymous referee for suggesting the use of more refined signal waveforms and providing constructive comments. We thank  S.~R.~Valluri and N. Arutkeerthi for helpful comments.
This work is supported by NSF Grant No.~PHY-2207935. We acknowledge the Texas Advanced Computing Center (TACC) at the University of Texas at Austin (www.tacc.utexas.edu) for providing high performance computing resources. This research has made use of data or software obtained from the Gravitational Wave Open Science Center (gwosc.org), a service of the LIGO Scientific Collaboration, the Virgo Collaboration, and KAGRA. 

\end{document}